\documentclass[12pt,preprint]{aastex}
\usepackage{graphicx} 
\newcommand{\ben}{\begin{enumerate}}
\newcommand{\een}{\end{enumerate}}
\newcommand{\bfig}{\begin{figure}}
\newcommand{\efig}{\end{figure}}
\newcommand{\beq}{\begin{equation}}
\newcommand{\eeq}{\end{equation}}
\newcommand{\mbf}{\mathbf}
\shorttitle{Minimum Structure Inductive Velocity Reconstruction}
\shortauthors{Georgoulis \& LaBonte}
\received{}
\begin{document}
\title{Reconstruction of an Inductive Velocity Field Vector 
from Doppler Motions and a Pair of Solar Vector Magnetograms}
\author{Manolis K. Georgoulis \& Barry J. LaBonte}
\affil{The Johns Hopkins University Applied Physics Laboratory,
        11100 Johns Hopkins Rd., Laurel, MD 20723, USA}
\begin{abstract}
We outline a general methodology to infer the inductive velocity field vector 
in solar active regions. For the first time, both the field-aligned and the 
cross-field velocity components are reconstructed. The cross-field velocity solution 
accounts for the changes of the vertical magnetic field seen between a pair of 
successive active-region vector magnetograms via the ideal induction equation. 
The field-aligned velocity is obtained using the Doppler velocity 
and the calculated cross-field velocity. Solving the ideal induction equation in 
vector magnetograms measured at a given altitude in the solar atmosphere is an 
under-determined problem. In response, our general formalism allows the use of 
any additional constraint for the inductive cross-field velocity to 
enforce a unique solution in the induction equation. As a result, our methodology 
can give rise to new velocity solutions besides the one presented here. 
To constrain the induction equation, we 
use a special case of the minimum structure approximation that 
was introduced in previous studies and is already employed here 
to resolve the $180^o$-ambiguity in the input vector  
magnetograms. We reconstruct the inductive velocity for 
three active regions, including NOAA AR 8210 for which previous results 
exist. Our solution believably reproduces the horizontal 
flow patterns in the studied active regions but breaks down 
in cases of localized rapid magnetic flux emergence or 
submergence. Alternative approximations and constraints are 
possible and can be accommodated into our general formalism. 
\end{abstract}
\keywords{MHD---Sun: atmospheric motions---Sun: magnetic fields---Sun: photosphere}
\section{Introduction}
Despite recent advances in vector magnetography, reliable 
measurement of the magnetic field vector in the solar 
active-region atmosphere can be 
routinely performed only at photospheric heights. In view of the central role 
of the magnetic field in the dynamical evolution and the eruptive manifestations 
in the solar atmosphere, the photospheric magnetic field vector should 
then be fully 
exploited to allow an assessment of the three-dimensional (3D) magnetic structures 
above the photosphere. Magnetic field extrapolations provide a means 
to calculate the chromospheric and coronal magnetic 
fields using the photospheric fields as the required boundary condition. 
Extrapolation techniques use approaches of various sophistication levels, 
such as the current-free approximation (Schmidt 1964), the linear force-free 
(Alissandrakis 1981), or the nonlinear force-free approximation 
(Amari, Boulmezaoud, \& Miki\'{c} 1999). Nonetheless, 
non-force-free extrapolations are probably the most realistic approach, given that 
the photospheric magnetic fields are forced (Metcalf et al. 1995;  
Georgoulis \& LaBonte 2004). The non-force-free modeling of the coronal magnetic 
fields requires the equations of Magnetohydrodynamics (MHD). 
Development of 3D MHD models has become possible due to the 
steadily increasing computing 
capabilities. These models are promising for understanding 
the pre-event configuration of solar eruptions and coronal mass ejections (CMEs), 
as well as the role of 
current sheets, turbulence, and kinematic effects in the 
solar atmosphere (see, e.g., Gudiksen \& Nordlund 2002; Amari et al. 2003). 
A particularly appealing aspect of 3D MHD models is that they can be data-driven, 
i.e., they can utilize boundary conditions corresponding to observed active region 
magnetic fields (Abbett 2003; Roussev et al. 2004). However, the electric fields, 
the magnetic forces, and the time-dependent parts of the MHD equations 
depend sensitively on the flows of the 
magnetized plasma. Therefore, photospheric boundary information should include a 
reliable assessment of the velocity field vector in addition to the 
magnetic field vector. 

The coupling between the photospheric magnetic and velocity fields also 
pertains to the calculation of the 
magnetic helicity in solar active regions (see Berger 1999 for an introduction). 
Helicity studies are important because they may imply a potential forecasting capability 
for solar flares and CMEs. Berger \& Field (1984) derived the Poynting theorem 
of magnetic helicity in an open volume. This expression 
enables the calculation of the temporal variation of the relative magnetic helicity 
above the boundary plane of the magnetic field measurements 
but employs explicitly the velocity field vector on the boundary plane.

Measurements of the photospheric magnetic field vector are routine but the 
photospheric velocity is only partially measured. Its longitudinal component 
can be inferred from the Doppler effect on the spectral line 
used for the magnetic field observations. However, the measurement of the Doppler 
velocity is subject to a number of caveats and ambiguities (see, e.g., Chae, Moon, \& 
Pevtsov 2004 for a discussion). In addition, the Doppler 
signal is affected by oscillatory effects, such as the $5-min$ oscillations, 
while its amplitude depends on the employed inference technique, with more than one 
such techniques existing. 
As a result, the Doppler velocity requires very careful handling. In addition, the 
Doppler velocity does not provide any information about the transverse 
(i.e. perpendicular to the line-of-sight) flows. 
Even if the problems associated with the Doppler velocity were resolved and 
even if the transverse velocity could be measured on the magnetically sensitive 
spectral line, however, 
knowledge of the photospheric flow field in solar active regions 
would still be partial. 
This is because the photospheric plasma is a mixture of magnetized and unmagnetized 
elements. Non-magnetized plasma flows have minimal impact on the magnetic field 
measurements and hence they cannot be inferred from the magnetically sensitive 
spectral line. This fact, however, may has a positive side since the 
knowledge of the purely hydrodynamic (HD) 
flows may not be crucial for the study of magnetic fields and their evolution. 

The first method used to infer the unknown transverse 
photospheric flows was the {\it local correlation tracking} (LCT) 
technique of November \& Simon (1988). The LCT algorithm measures 
the displacements of the various patterns 
present in a pair of white-light continuum 
images and finds an average velocity associated with these displacements 
under the assumption that the motions causing the displacements are smooth.
A major requirement for the LCT method to work is that the studied patterns must 
exhibit significant contrast differences to be followed effectively in time. Moreover, 
the patterns must maintain their structural integrity for time scales much larger 
than the cadence of the observations since a continuous 
restructuring would make tracking problematic. The above two requirements 
are fulfilled  in the quiet photospheric filigree observed in optical 
wavelengths, so the LCT technique is undoubtedly the best way to 
infer the transverse HD flows of the mostly unmagnetized quiet solar photosphere. 
November \& Simon (1988) showed that the LCT velocity successfully reproduces the 
quiet-Sun granular convective motion.  
Tracking algorithms have been implemented in a number of schemes, such as the 
``feature tracking'' method of Strous et al. (1996), the LCT technique 
of  Berger et al. (1998), and the ``balltracking'' technique of Potts, Barrett, 
\& Diver (2004). 

Despite its importance in quiet-Sun velocity measurements, the LCT technique 
is problematic in the active-region photosphere. 
First, it is not clear which input would 
give more appropriate results to LCT algorithms, as both 
the continuum images and the magnetogram images reveal different aspects 
of the photospheric flows. Applying the two different inputs 
leads to two different results. Second, both the white-light 
and the magnetogram structures in active regions may reform 
too rapidly for flows to be 
inferred with sufficient detail. The 
photospheric filigree is present but disrupted because of the 
widespread strong magnetic flux, 
while rapid restructuring of the emerging magnetic structures, 
especially in emerging flux regions, introduces uncertainties in the 
LCT results. Consequently, it is not uncommon for the LCT velocity 
to require significant averaging in order 
to yield results consistent with a visual inspection of the 
observed flows in active regions. 
Even averaging needs to be performed carefully to avoid 
propagation of systematic errors (Roudier et al. 1999). 
Third, the LCT technique requires a FWHM apodizing window 
whose size is typically a few times larger than the instrument's pixel 
size. Therefore, the LCT flow maps are coarser than the employed input maps. 
This is because the LCT maximizes the cross-correlation function by applying 
displacements to the initial tiled image and 
by comparing the result with the second 
image of the pair, so each apodization 
tile must contain sufficient structure to yield a well-defined 
peak of the cross-correlation function. Different 
choices of the apodizing window reflect different types of flows and give 
different results, since the kinetic power is distributed over a variety of spatial 
scales in an active region. Small tile sizes tend to reveal the small-scale 
convection while large tile sizes reproduce the systematic flows seen during 
the emergence of the magnetic dipole(s) that eventually form an active region. 
Fourth, some systematic motions do not even correspond to actual flows, but they are 
apparent and caused by the projected emergence of inclined magnetic structures. 
Tracking techniques tend to attribute a systematic transverse velocity to these 
purely vertical flows. In summary, as an active region traverses far from disk 
center, the LCT velocity provides a crude, as well as inextricable, 
mixture of HD flows, MHD flows, horizontal and vertical flows, 
and apparent flows caused by projection effects. 

Given that the velocity field acting on a given magnetic field is required to 
calculate the magnetic helicity flux on the plane of the observations, several 
authors  (e.g., Moon et al. 2002; Nindos \& Zhang 2002; 
Romano, Contarino, \& Zuccarello 2003) 
have used the LCT velocity inferred by photospheric magnetograms after 
Chae (2001) developed a technique to evaluate the Poynting theorem of Berger \& 
Field (1984). 
In these studies, the LCT velocity is assumed representative of the 
actual horizontal velocity which is inaccurate even at moderate distances from 
disk center. Moreover, the impact of the apparent transverse flows caused by 
projection effects is not removed. D\'{e}moulin \& Berger (2003) quantified this effect 
by showing that the LCT velocity 
$\mbf{v_{LCT}}$ relates to both the actual vertical velocity $u_z$ and the actual 
horizontal velocity $\mbf{u_h}$ via a relation of the form
\beq
\mbf{u_h}={{u_z} \over {B_z}} \mbf{B_h} + \mbf{v_{LCT}}\;\;,
\label{db}
\eeq
where $B_z$ and $\mbf{B_h}$ are the vertical magnetic field and the horizontal magnetic 
field vector, respectively (considering the spherical geometry of the Sun, 
these components can be equivalently called the normal and the tangential 
to the photosphere components, respectively). 
From equation (\ref{db}) we notice that the LCT velocity can be 
very different from the actual horizontal velocity $\mbf{u_h}$ in the presence of 
large magnetic fluxes and emerging or submerging inclined magnetic configurations. 

Since MHD simulations and magnetic helicity calculations require a velocity field 
vector that reflects the evolution of the magnetic field and the magnetized 
plasma flows, an alternative way to calculate the velocity is to demand minimum 
compliance with the MHD equations. A fundamental time-dependent MHD equation is 
the induction equation 
\beq
{{\partial \mbf{B}} \over {\partial t}} = \mbf{\nabla} \times (\mbf{u} \times \mbf {B})
+ \eta \nabla ^2 \mbf{B}\;\;,
\label{vin}
\eeq
where the action of the velocity field $\mbf{u}$ on the magnetic field $\mbf{B}$ 
together with the diffusion of the magnetic field subject to a magnetic diffusivity 
$\eta$ are responsible for the temporal variation of the magnetic field. To solve 
equation (\ref{vin}) for $\mbf{u}$ it is necessary to adhere to the ideal limit 
of the induction equation, i.e. to assume that the diffusive term 
$\eta \nabla ^2 \mbf{B}$ is negligible. 
Furthermore, single-height (i.e. photospheric) vector magnetic field measurements 
preclude direct knowledge of the vertical gradient $(\partial / \partial z)$ of any 
measured quantity. As a result, we can only tackle 
the vertical component of the ideal MHD induction equation, namely, 
\beq
{{\partial B_z} \over {\partial t}} = 
[\mbf{\nabla} \times (\mbf{u} \times \mbf {B})]_z\;\;.
\label{iz}
\eeq
The goal, then, is to solve equation (\ref{iz}) for $\mbf{u}$ using a pair of 
vector magnetograms to calculate $(\partial B_z / \partial t)$. A major drawback 
to this objective is that the problem is under-determined: the induction equation 
(\ref{iz}) by itself does not have a unique solution. Evidently, only the cross-field 
(i.e. perpendicular to the magnetic field vector) velocity $\mbf{u_{\perp}}$ 
participates in the induction equation. This contributes a condition additional 
to the ideal induction equation (\ref{iz}), namely the orthogonality of the 
$\mbf{u_{\perp}}$-solution with $\mbf{B}$, $\mbf{u_{\perp}} \cdot \mbf{B}=0$.
As a result, only two equations are available 
to infer the three components of $\mbf{u_{\perp}}$. An additional assumption or 
constraint may lead to a third equation and hence to a unique solution of equation 
(\ref{iz}). Since only $\mbf{u_{\perp}}$ participates in the induction equation, 
the field-aligned velocity $\mbf{u_{||}}$ can only be inferred 
by additional assumptions, independent of the induction equation. 

Kusano et al. (2002) first devised a technique to constrain the induction equation 
(\ref{iz}) by utilizing the LCT velocity. 
The latter was assumed representative of the actual horizontal velocity and then the 
induction equation (\ref{iz}) was solved for the vertical velocity. 
Next, Welsch et al. (2004) explicitly employed the geometrical expression of 
D\'{e}moulin \& Berger (equation (\ref{db}) and $\mbf{v_{LCT}}$. 
By means of the orthogonality condition, $\mbf{u_{\perp}} \cdot \mbf{B}=0$, 
equation (\ref{db}) and equation (\ref{iz}), then, they were able to obtain 
a unique [called the inductive LCT (ILCT)] solution for $\mbf{u_{\perp}}$. 

The ``hybrid'' techniques of Kusano et al. (2002) 
and Welsch et al. (2004) are certainly advancements over a simple use of 
the LCT velocity but but they give rise to a number of issues:
\ben
\item[(1)] The LCT velocity may be inconsistent with the induction equation, 
since (i) MHD is not used to infer the LCT velocity, (ii) the 
LCT velocity reflects a mixture of HD, MHD, 
and apparent flows, measured on a coarser scale than the corresponding 
vector magnetograms, and (iii) the LCT results depend on the choice of 
the FWHM apodizing window. 
\item[(2)] The use of the LCT velocity may enforce a unique solution of the 
induction equation, but the LCT velocity itself is {\it not} unique. 
A different LCT result gives rise to a different inductive velocity 
$\mbf{u_{\perp}}$. How sensitively $\mbf{u_{\perp}}$ depends $\mbf{v_{LCT}}$ 
is unknown. 
\een
 
The first inductive technique for a $\mbf{u_{\perp}}$ disentangled from the LCT 
velocity was developed by Longcope (2004) who found a parametric family of  
solutions of the induction equation (\ref{iz}) and then selected the solution 
with minimal flow speed. Longcope's (2004) 
minimum energy fit (MEF) technique is an excellent development.  
The MEF hypothesis is reasonable, appropriate for helicity studies, 
and most probably consistent with 3D MHD models. Nevertheless, 
it is not clear whether 
the complicated evolution of complex, dynamic active regions is consistent with 
an assumption of minimum flows. 

Notice that all the above inductive techniques consider only the cross-field 
velocity $\mbf{u_{\perp}}$ and make no attempt to calculate the field-aligned 
velocity $\mbf{u_{||}}$. 

In this study, we generalize the reconstruction of an inductive 
velocity field vector in solar active regions. 
We attempt a complete reconstruction including {\it both} 
the field-aligned and the cross-field velocity components that account for the 
change of the vertical magnetic field seen between a pair of observed vector 
magnetograms. First, we outline a general solving technique that allows 
{\it any} additional constraint for $\mbf{u_{\perp}}$ 
to enforce a unique solution of the induction equation 
(\ref{iz}) for the cross-field velocity. Second, we calculate 
the field-aligned velocity by rotating the Doppler (longitudinal) velocity 
to the heliographic reference system. Like Longcope (2004) we refrain from 
using the LCT velocity. 
Our proposed constraint is a restricted {\it minimum structure} approximation. 
The unrestricted approximation was introduced 
by Georgoulis, LaBonte, \& Metcalf (2004) to resolve the $180^o$-ambiguity in the 
orientation of the transverse magnetic field component in vector magnetogram 
measurements and is already employed here since we cannot 
apply the induction equation to a pair of $180^o$-ambiguous vector magnetograms. 
The restriction in the minimum structure approximation stems from assuming 
that the magnetic field $\mbf{B}$, the cross-field velocity $\mbf{u_{\perp}}$, and 
the cross-field gradient $(\mbf{\nabla}B)_{\perp}$ of the magnetic field strength 
are {\it coplanar}. This assumption implies that 
$\mbf{u_{\perp}}$ directly relates to the gradients $(\nabla B)_{\perp}$ 
and aims to either sustain or relax them on the 
$[(\mbf{\nabla}B)_{\perp},\mbf{B}]$-plane. This ``coplanar minimum structure 
approximation'' enforces a unique solution of the induction equation (\ref{iz}) for 
$\mbf{u_{\perp}}$. Together with the corresponding field-aligned velocity 
$\mbf{u_{||}}$ we perform a complete 
{\it minimum structure reconstruction} (MSR) of the 
inductive velocity field. The general solution of the 
ideal induction equation (\ref{iz}) and the general calculation of the field-aligned 
velocity are discussed in \S2. The analytical 
derivation of a velocity field consistent with the MHD equations 
is performed in \S3. The special case of the MSR velocity solution is derived 
and discussed in \S4. The MSR velocity for three solar active regions 
is reconstructed in \S5, while in \S6 we discuss our physical assumptions and we 
summarize our findings. 
\section{General reconstruction of the inductive velocity field vector}
\subsection{Cross-field velocity: a general solution of the induction equation}
A useful technique to treat the vertical component of the ideal induction 
equation has been described by Welsch et al. (2004) and Longcope (2004). 
The rhs of equation (\ref{iz}) can be transformed into the divergence of a 
vector as follows: 
\beq
{{\partial B_z} \over {\partial t}}= \mbf{\nabla}_h \cdot 
(v_z \mbf{B_h} - B_z \mbf{v_h})\;\;,
\label{lr1}
\eeq
where $\nabla _h=[(\partial / \partial x), (\partial / \partial y), 0]$.
Then, by introducing two scalar potentials 
$\phi$ and $\psi$, termed a ``stream function'' and a ``electrostatic potential'', 
respectively, by Longcope (2004), the term in the 
parenthesis in equation (\ref{lr1}) can be written as 
\beq
v_z \mbf{B_h} - B_z \mbf{v_h} = \mbf{\nabla}_h \phi + 
\mbf{\nabla}_h \psi \times \mbf{\hat{z}}\;\;.
\label{lr2}
\eeq
To find a unique solution $\mbf{u_{\perp}}$ 
for the inductive velocity $\mbf{v}$, then, one has to 
specify uniquely the scalar potentials $\phi$ and $\psi$. 
\subsubsection{Calculating the stream function $\phi$}
There are at least two ways to find a unique solution for the stream function $\phi$.
The simplest way is to 
substitute equation (\ref{lr2}) into equation (\ref{lr1}). This 
yields a Poisson equation for $\phi$ on the horizontal plane, namely, 
\beq
{{\partial B_z} \over {\partial t}}= \nabla _h^2 \phi\;\;.
\label{pot}
\eeq
Equation (\ref{pot}) can be solved by means of a 
{\it successive over-relaxation} (SOR) technique or 
an {\it alternating-direction implicit} (ADI) technique 
(\S2.3 and Press et al. 1992). The SOR algorithm 
is simpler, albeit computationally 
expensive for large grids. Equation (\ref{pot}) can be solved for $\phi$ 
on the horizontal plane $S$ assuming 
homogeneous Dirichlet boundary conditions ($\phi =0$) on $\partial S$. 
The calculation of $\phi$ alone yields a special inductive velocity 
$\mbf{v}$ given by 
\begin{eqnarray}
\begin{array}{l}
\mbf{v_h}=(1/B_z) (- \nabla _h \phi + v_z \mbf{B_h})\\
v_z=(1/B^2) \mbf{B_h} \cdot \nabla _h \phi\;\;.\\
\end{array}
\label{nus2}
\end{eqnarray}
To reach the special solution $\mbf{v}$ we used the orthogonality condition 
$\mbf{v} \cdot \mbf{B}=0$ in conjunction with equation (\ref{lr2}) for $\psi=0$.

Alternatively, one may write equation (\ref{iz}) as a flux-conservative equation 
on the horizontal plane, namely, 
\beq
\mbf{\nabla _h} \cdot \mbf{F} = {{\partial B_z} \over {\partial t}}\;\;,
\label{fc}
\eeq
where $\mbf{F}=[(\mbf{v} \times \mbf{B})_y,-(\mbf{v} \times \mbf{B})_x,0]$ 
is the conserved ``flux'' and $\mbf{v}$ is the inductive 
velocity. If we assume that $\mbf{v}$ is identical to the velocity of 
equations (\ref{nus2}), obtained for $\psi =0$, then 
the flux $\mbf{F}=\mbf{\nabla}_h \phi$, and hence the stream function $\phi$, can be 
calculated by means of a {\it timestep splitting} technique 
(Boris et al. 1993). The latter approach requires less computing time compared to 
the SOR and similar computing time to the ADI algorithm. 
\subsubsection{Calculating the electrostatic potential $\psi$}
In previous inductive techniques the additional constraint used to enforce a 
unique solution of the ideal induction equation has been either the LCT velocity 
(Kusano et al. 2002; Welsch et al. 2004) or an assumption of minimal flows 
for the desired flow field (Longcope 2004). We hereby generalize the problem 
by uniquely specifying the electrostatic potential $\psi$ when any one component 
of the desired solution $\mbf{u_{\perp}}$ is assumed known on dependent on the 
other two components of $\mbf{u_{\perp}}$. Without loss of generality we choose 
the vertical component $u_{\perp _z}$ of $\mbf{u_{\perp}}$ as 
the prescribed component. 
Besides leading to a unique solution for $\mbf{u_{\perp}}$, 
this action can reproduce the results of previous inductive techniques 
if these techniques' $u_{\perp _z}$-solution is introduced into our 
general solving methodology or it can give rise to new solutions if 
alternative prescriptions for $u_{\perp _z}$ are used. 

To find a unique solution for $\psi$, 
recall the flux-conservative equation (\ref{fc}), where 
$\mbf{F}=[(\mbf{v} \times \mbf{B})_y,-(\mbf{v} \times \mbf{B})_x,0]$ 
is the flux and $\mbf{v}$ is the special inductive velocity of equations 
(\ref{nus2}). We realize that any flux 
$\mbf{F'}=[(\mbf{v'} \times \mbf{B})_y,-(\mbf{v'} \times \mbf{B})_x,0]$ is a 
solution of equation ({\ref{fc}}), and hence of the induction equation (\ref{iz}), 
if it satisfies the condition
\beq
\mbf{F'}=\mbf{F} + \mbf{G}\;\;;\;\;\;with\;\;\;
\mbf{\nabla}_h \cdot \mbf{G}=0\;\;.
\label{fp}
\eeq
Notice that $F_z=F'_z=0$, so $G_z$=0. Therefore, one may define two gauge conditions 
for $\mbf{G}$ in the volume $z \ge 0$ and on the plane $S$ ($z=0$) 
of the magnetic field measurements, namely 
\beq
\mbf{\nabla} \cdot \mbf{G}=0\;\;,\;\;\;and\;\;\;
\mbf{G} \cdot \mbf{\hat{z}} |_{z=0}=0\;\;.
\label{gg}
\eeq
Combining equations (\ref{lr2}), (\ref{fc}), and (\ref{fp}) one finds 
that $\mbf{G}$ is related to the electrostatic potential $\psi$ by the relation 
$\mbf{G}=\mbf{\nabla}_h \psi \times \mbf{\hat{z}}$. Therefore $\mbf{G}$ can be 
named an ``electrostatic field''. To find a unique solution for the induction 
equation (\ref{iz}), therefore, one needs a unique solution for 
the electrostatic potential $\psi$ or, 
equivalently, a unique solution for the electrostatic field $\mbf{G}$. 

Given the gauge conditions of equation (\ref{gg}), calculating $\mbf{G}$ is 
equivalent to the vector potential calculation of Chae (2001) and it can be 
performed in Fourier space. Indeed, the solution for 
the electrostatic field $\mbf{G}$ reads (see also Chae 2001)
\begin{eqnarray}
\begin{array}{l}
G_x = \mathcal{F}^{-1} [(ik_y/(k_x^2+k_y^2)) \mathcal{F}(C_z)]\\
G_y = \mathcal{F}^{-1} [-(ik_x/(k_x^2+k_y^2)) \mathcal{F}(C_z)]\;\;,\\
\end{array}
\label{Gs}
\end{eqnarray}
where $\mathcal{F}(r)$, $\mathcal{F}^{-1}(r)$ are the direct and the inverse Fourier 
transform of $r$, respectively, and 
\beq
C_z=(\mbf{\nabla} \times \mbf{G})_z\;\;.
\label{cz}
\eeq
However, a major difference between our problem and Chae's (2001) is that $C_z$ is 
unknown and hence equations (\ref{Gs}) cannot be evaluated. If $u_{\perp _z}$ is 
known, however, $\mbf{G}$ and 
$C_z$ can be specified iteratively and self-consistently, in accordance with the 
gauge conditions of equations (\ref{gg}). In essence, we iteratively 
solve the system of the three equations (\ref{gg}) and (\ref{cz}) for the two 
unknown horizontal components of $\mbf{G}$ and for $C_z$ by using equations 
(\ref{Gs}). The system is closed and 
has a unique solution for $\mbf{G}$ and $C_z$ and, by extension, a unique solution 
for $\psi$. 

To illustrate how knowledge of $u_{\perp _z}$ is sufficient to calculate $\mbf{G}$ 
and $C_z$, let 
$\mbf{u_{\perp}}$ be the desired unique solution of the induction equation 
(\ref{iz}) and $u_{\perp _z}$ be its known vertical component. Then, from equation 
(\ref{lr2}) one obtains 
\beq
\mbf{u_{\perp _h}}={{1} \over {B_z}} (u_{\perp _z} \mbf{B_h} - \nabla _h \phi - \mbf{G})\;\;,
\label{fsol}
\eeq
where we have introduced the definition 
$\mbf{G}=\mbf{\nabla}_h \psi \times \mbf{\hat{z}}$ in equation (\ref{lr2}).
Solving equation (\ref{fsol}) for $\mbf{G}$ and substituting 
$(\partial \phi / \partial x)$, $(\partial \phi / \partial y)$ 
with the components of $\mbf{v}$ from equations 
(\ref{nus2}), one further obtains 
\beq
\mbf{G}=(u_{\perp _z} -v_z) \mbf{B} + \mbf{G_{\perp}}\;\;,
\label{it1}
\eeq
where $\mbf{G_{\perp}}=-B_z(\mbf{u_\perp} - \mbf{v})$. 
Obviously, $\mbf{G_{\perp}}$ is a purely cross-field electrostatic field 
since $\mbf{u_{\perp}} \perp \mbf{B}$ 
and $\mbf{v} \perp \mbf{B}$. The field-aligned 
component of $\mbf{G}$ in equation (\ref{it1}) 
is fully known since $u_{\perp _z}$, $v_z$, and $\mbf{B}$ are known. If 
$\mbf{G_{\perp}}$ is not accurately known, however, i.e. if 
$\mbf{G_{\perp}} \cdot \mbf{B} \ne 0$, then the associated 
error is given by the quantity 
$\Delta E= \mbf{G_{\perp}} \cdot \mbf{B}$ or, equivalently, 
\beq
\Delta E=\mbf{G} \cdot \mbf{B} - (u_{\perp_z} - v_z)B^2\;\;.
\label{it2}
\eeq
One may then introduce a correction $\mbf{\delta G}$ to the electrostatic field 
$\mbf{G}$, namely, 
\beq
\mbf{G'}=\mbf{G}+\mbf{\delta G}\;\;,
\label{it3}
\eeq
such that $\mbf{G' _{\perp}} \cdot \mbf{B}=0$, where 
$\mbf{G'_{\perp}}=\mbf{G'}- (u_{\perp _z} -v_z) \mbf{B}$. This obviously implies 
that $\mbf{\delta G} \cdot \mbf{B}=- \Delta E$. A general form for $\mbf{\delta G}$ 
would then be
\beq
\mbf{\delta G}= - \Delta E ({{1} \over {k_1 B_x}}, {{1} \over {k_2 B_y}},0)\;\;;
\;\;\;\;{{1} \over {k_1}}+{{1} \over {k_2}}=1\;\;,
\label{it4a}
\eeq
where $k_1$, $k_2$ are real numbers. Assuming that $\mbf{\delta G}$ has the minimum 
possible magnitude so that the correction from $\mbf{G}$ to $\mbf{G'}$ is the 
minimum possible, we can calculate $k_1$, $k_2$ and express $\mbf{\delta G}$ as 
follows:
\beq
\mbf{\delta G}=-{{\Delta E} \over {B_h^2}}(B_x,B_y,0)\;\;.
\label{it4}
\eeq
Based on the above considerations, we can now calculate iteratively an electrostatic 
field $\mbf{G}$ that satisfies both the gauge conditions of equation (\ref{gg}) 
and the orthogonality condition $\mbf{G_{\perp}} \cdot \mbf{B}=0$ for its 
cross-field component $\mbf{G_{\perp}}$. The devised iterative scheme is as follows:
\ben
\item[(I)] At iteration 0, use 
$\mbf{G}^{(0)}= (u_{\perp _z}-v_z)\mbf{B_h} + \mbf{v_h}B_z$ as a first 
guess for $\mbf{G}$\newline
(from equation (\ref{it1}) assuming than $\mbf{u_{\perp _h}}=0$ in $\mbf{G_{\perp}}$)

..................................................................................................

\item[(II)] At iteration $n$ ($n=0,1,2,...$), calculate 
$C_z^{(n)}=(\mbf{\nabla} \times \mbf{G}^{(n)})_z$ 
\item[(III)] At iteration $n+1/2$, evaluate 
equations (\ref{Gs}) for  $\mbf{G}^{(n+1/2)}$\newline 
(thus solving equation (\ref{cz}) while enforcing the gauge conditions of equation (\ref{gg}))
\item[(IV)] At iteration $n+1/2$, calculate the correction $\mbf{\delta G}^{(n+1/2)}$ of 
$\mbf{G}^{(n+1/2)}$ from  equations (\ref{it2}) and (\ref{it4})
\item[(V)] At iteration $n+1$, update 
$\mbf{G}^{(n+1)} =  \mbf{G}^{(n+1/2)} + \mbf{\delta G}^{(n+1/2)}$ and continue from 
step II, etc. 
\een
Because we demand that $\mbf{\delta G}$ has the minimum possible 
magnitude, $\delta G = |\Delta E|$, which decreases with each iteration, 
the above scheme converges asymptotically to a solution for the electrostatic field 
$\mbf{G}$ with correction $\mbf{\delta G} \rightarrow 0$. Convergence at iteration 
$n$ is checked by means of a normalized dimensionless ratio $\mathcal{R}$ of the form
\beq
\mathcal{R} = {{\sum ( |G_x^{(n)} - G_x^{(n-1)}| + |G_y^{(n)} - G_y^{(n-1)}| )} 
\over {\sum ( |G_x^{(n)}| + |G_x^{(n-1)}| + |G_y^{(n)}| + |G_y^{(n-1)}| )}}\;\;,
\label{c1}
\eeq
or, alternatively, 
\beq
\mathcal{R} = {{\sum ( |\mbf{G}^{(n)}-  \mbf{G}^{(n-1)}| )} \over
{\sum ( |\mbf{G}^{(n)}|+  |\mbf{G}^{(n-1)}| )}}\;\;,
\label{c2}
\eeq
where the summation includes only locations with well-measured magnetic flux, i.e., 
locations where the vertical and the horizontal magnetic field components 
are larger than their associated uncertainties. Because we use discrete Fourier 
transforms the above calculations are somewhat approximate, so 
$\mathcal{R} =0$ is achieved 
with respect to a prescribed fractional tolerance limit $\varepsilon _c$. 
Either form of $\mathcal{R}$ (equations (\ref{c1}), (\ref{c2}))
leads to the same results. The process stops at iteration $n$ if the fraction 
\beq
\varepsilon={{|\mathcal{R}^{(n)}-\mathcal{R}^{(n-1)}|} \over 
{\mathcal{R}^{(n)}+\mathcal{R}^{(n-1)}}} < \varepsilon _c\;\;.
\label{epsi}
\eeq
In the examples of \S5 we have used $\varepsilon _c = 10^{-4}$. 

After its calculation, we can use the electrostatic field $\mbf{G}$ in equation 
(\ref{fsol}) to calculate the horizontal inductive cross-field 
velocity $\mbf{u_{\perp _h}}$. 
The vertical component $u_{\perp _z}$ is {\it a priori} 
known, so one obtains a unique solution of the induction equation (\ref{iz}) for the 
cross-field velocity $\mbf{u_{\perp}}$. Notice that the above formalism 
to infer $\mbf{G}$ can be 
modified accordingly if $u_{\perp _x}$ or $u_{\perp _y}$, rather than $u_{\perp _z}$, 
are assumed known.  
\subsection{Field-aligned velocity: rotation of the Doppler velocity}
Besides the cross-field velocity $\mbf{u_{\perp}}$ one also needs the field-aligned 
velocity $\mbf{u_{||}}$ to fully reconstruct the inductive velocity field 
$\mbf{u}=\mbf{u_{\perp}}+\mbf{u_{||}}$. The 
field-aligned velocity cannot be treated by the induction equation and therefore 
its calculation requires additional information or assumptions. In the formulation 
of \S2.1 the vertical cross-field velocity $u_{\perp _z}$ is 
assumed known but the vertical 
field-aligned velocity $u_{|| _z}$ and the total vertical velocity 
$u_z=u_{|| _z} + u_{\perp _z}$ are unknown. Evidently, if either 
$u_{|| _z}$ or $u_z$ are known together with $u_{\perp _z}$, the field-aligned 
velocity $\mbf{u_{||}}$ can be calculated by the relation 
\beq
\mbf{u_{||}}={{u_{|| _z}} \over {B_z}} \mbf{B}=
{{u_z - u_{\perp _z}} \over {B_z}} \mbf{B}\;\;.
\label{ufa1}
\eeq
Therefore, the problem of calculating the field-aligned flows is essentially a 
problem of calculating $u_{|| _z}$ or $u_z$ for a given  $u_{\perp _z}$. To 
calculate $u_z$ we use the longitudinal Doppler velocity $u_l$, 
available for nearly any 
vector magnetogram measurement of a given active region. From the geometrical 
transformation to the heliographic reference system (Gary \& Hagyard 1990) 
the longitudinal velocity $u_l$ and the heliographic components $u_x$, $u_y$, $u_z$ 
of the velocity field vector $\mbf{u}$ on the observer's (image) plane are related 
as follows:
\beq
u_l = \alpha u_x + \beta u_y + \gamma u_z\;\;,
\label{dc}
\eeq
where $\alpha$, $\beta$, $\gamma$ are the direction cosines of the transformation. 
Their values can be calculated at each location of the image plane by means of 
this location's heliographic latitude and longitude, the heliographic latitude and 
longitude of the center of the solar disk, and the angular position of the northern 
extremity of the solar rotation axis, measured eastward from the northernmost point 
of the solar disk (solar $P$-angle). 
Decomposing the velocity components $u_i$; $i \equiv \{ x,y,z \}$, 
into a field-aligned and a cross-field terms, $u_{|| _i}$ and $u_{\perp _i}$, 
respectively, equation (\ref{dc}) can be solved for $u_z$ to give 
\beq
u_z={{B_z} \over {B_l}} (u_l - \alpha u_{\perp _x} - \beta u_{\perp _y} -\gamma 
u_{\perp _z}) + u_{\perp _z}\;\;,
\label{uz}
\eeq
where $B_l = \alpha B_x + \beta B_y + \gamma B_z$ is the longitudinal magnetic 
field related to the heliographic magnetic field vector $(B_x,B_y,B_z)$ 
by means of the direction cosines $\alpha$, $\beta$, and $\gamma$. The first term 
in the rhs of 
equation ({\ref{uz}}) obviously corresponds to the field-aligned vertical velocity 
$u_{|| _z}$. Equation (\ref{uz}) suggests that the Doppler velocity $u_l$ alone 
does not suffice for the calculation of the vertical velocity $u_z$ and hence of 
the field-aligned flows. One also needs the cross-field velocity 
$\mbf{u_{\perp}}$ obtained after solving the induction equation. 
On disk center, equation (\ref{uz}) directly implies $u_z=u_l$ 
since $B_z=B_l$ and $\alpha = \beta = 0$, while $\gamma =1$. 

Summarizing, we describe generally 
how an inductive velocity field $\mbf{u}$ can be fully 
reconstructed from its field-aligned and cross-field components. The cross-field 
velocity $\mbf{u_{\perp}}$ is obtained as a solution of the ideal induction 
equation, while the field-aligned velocity $\mbf{u_{||}}$ is calculated as 
shown in equation (\ref{uz}) using the Doppler velocity and 
the inferred cross-field 
velocity to obtain the vertical component $u_z$ of $\mbf{u}$ or the vertical 
component $u_{|| _z}$ of $\mbf{u_{||}}$. 
\subsection{Algorithmic implementation}
Given a pair of co-aligned vector magnetograms, $\mbf{B_1}$ and $\mbf{B_2}$, obtained 
at times $t$ and $t+\Delta t$, respectively, we calculate the inductive 
cross-field velocity $\mbf{u_{\perp}}$ and the field-aligned velocity $\mbf{u_{||}}$ 
at time $t+(\Delta t/2)$ assuming that this velocity solution accounts for 
the evolution of the vertical magnetic field observed between $t$ and $t+\Delta t$. 
The velocity solution corresponds to an average magnetic field 
$\mbf{B}=(1/2)(\mbf{B_1}+\mbf{B_2})$ and an average Doppler velocity 
$u_l=(1/2)(u_{l_1}+u_{l_2})$, where $u_{l_1}$, $u_{l_2}$ are the Doppler velocities 
corresponding to the magnetograms $\mbf{B_1}$, $\mbf{B_2}$, respectively. 
The above averages are assumed to describe the magnetic configuration at time 
$t+(\Delta t/2)$. From these 
assumptions the temporal derivative $(\partial B_z / \partial t)$ in the 
induction equation (\ref{iz}) can be approximated by a finite difference 
between $B_{1_z}$ and $B_{2_z}$, namely,
\beq
{{\Delta B_z} \over {\Delta t}} \simeq {{1} \over {\Delta t}} 
(B_{2_z} - B_{1_z})\;\;.
\label{dBz_t}
\eeq
The horizontal derivatives $(\partial Q / \partial x)$, $(\partial Q / \partial y)$ 
of any quantity $Q$ can also be approximated by central finite differences of the form 
\begin{eqnarray}
\begin{array}{l}
(\Delta Q / \Delta x)=[1/ (2 \lambda)] [Q(x+1,y)-Q(x-1,y)]\\
(\Delta Q / \Delta y)=[1/ (2 \lambda)] [Q(x,y+1)-Q(x,y-1)]\;\;,\\
\end{array}
\label{cfd}
\end{eqnarray}
respectively, where $\lambda$ is the linear size of the magnetogram's pixel. 

As mentioned in \S2.1.1, Poisson's equation (\ref{pot}) is solved for the stream 
function $\phi$ on the plane $S$ of the magnetic 
field measurements assuming homogeneous Dirichlet boundary conditions equal to zero 
on $\partial S$. If the SOR algorithm is applied for this purpose then one also needs 
a spectral radius $\rho _{jac}=(1/2)[cos(\pi/L_1) + cos(\pi/L_2)]$ 
for the Jacobi relaxation, where a magnetogram with linear dimensions 
$L_1 \times L_2$ and a square pixel ($\lambda _x=\lambda _y=\lambda$) 
is assumed. This selection is a special case of the general 
examples discussed by Press et al. (1992). To further expedite the SOR, the 
Chebyshev acceleration scheme may be 
used in the selection of the over-relaxation parameter 
$\omega$, as also described by Press et al. (1992). In this study we use the faster, 
but more complicated, ADI algorithm. Since the ADI numerical scheme was originally 
introduced to solve the time-dependent ``heat flow'' equation 
$(\partial \phi / \partial t) - \eta \nabla ^2 \phi = \rho$ on a given plane, we use 
$\eta =1$, a fixed $\rho=-(\partial B_z / \partial t)$, and $t \rightarrow \infty$ 
to reach Poisson's equation (\ref{pot}). 

Both the SOR and the ADI algorithms can solve equation (\ref{pot}) 
with an accuracy restricted only by the machine accuracy 
for any given $(\Delta B_z / \Delta t)$. In the ILCT and MEF solutions 
both $\phi$ and $\psi$ are obtained by solving Poisson's equations, so any 
given temporal variation $(\Delta B_z / \Delta t)$ is reproduced up to machine 
accuracy for non-zero $\psi$. 
This is not the case for our general technique, where the electrostatic 
field $\mbf{G}$, and hence the electrostatic potential $\psi$, are calculated within 
a reasonable error. The error stems from the convergence process in minimizing the 
dimensionless ratio $\mathcal{R}$ (equations (\ref{c1}), (\ref{c2})) because of the 
use of the approximate Fourier transforms. Nevertheless, this error is not 
large enough to inflict a serious impact in the calculation of $\mbf{u_{\perp}}$. 
As we will see in the examples of \S5 the error in the minimization of $\mathcal{R}$ 
is smaller than $\sim 7$\% in all cases. Numerical tests have shown that the error 
decreases substantially when the quality and continuity 
of the magnetic field measurements improves, 
i.e., when the noise decreases and the smoothness of the magnetic field structure 
increases. For tested synthetic magnetograms 
free of observational noise (Abbett 2005, private communication), for example, 
the error in $\mathcal{R} \simeq 0$ is typically less than $1$\%.
\section{Analytical expressions for the velocity field}
In this Section we provide general analytical expressions for the velocity field 
$\mbf{u}$ acting on a given magnetic configuration $\mbf{B}$. In connection 
with the previous sections, we assume that $\mbf{B}$ is the average of two magnetic 
configurations, $\mbf{B_1}$ and $\mbf{B_2}$, obtained at times $t$ and 
$t + \Delta t$, respectively. We will derive both 
the field-aligned and the cross-field components of $\mbf{u}$. 

We begin by considering Amp\'{e}re's law in conjunction with the simplified Ohm's 
law for the electric current density $\mbf{J}$: from Amp\'{e}re's law, 
$\mbf{J}=(c/4 \pi) \mbf{\nabla} \times \mbf{B}$, replace $\mbf{B}$ by $B\mbf{\hat{b}}$ 
to obtain 
\beq
\mbf{J}=\mbf{J_1} + \mbf{J_2}\;,\;\;\;where\;\;\;
\mbf{J_1}={{cB} \over {4 \pi}} \mbf{\nabla} \times \mbf{\hat{b}}\;\;\;and\;\;\;
\mbf{J_2}={{c} \over {4 \pi}} \mbf{\nabla}B \times \mbf{\hat{b}}\;\;,
\label{amp}
\eeq 
where $\mbf{\hat{b}}$ is the unit vector along the magnetic field lines. From 
equation (\ref{amp}) we notice that $\mbf{J_2}$ is a purely cross-field component 
of the current density $\mbf{J}$ 
while $\mbf{J_1}$ includes, in general, a field-aligned and a cross-field component. 

From the simplified Ohm's law, on the other hand, one obtains 
\beq
\mbf{J}=\mbf{J'_1} + \mbf{J'_2}\;,\;\;\;where\;\;\;
\mbf{J'_1}= \sigma \mbf{E}\;\;\;and\;\;\;
\mbf{J'_2}= {{\sigma} \over {c}} \mbf{u} \times \mbf{B}\;\;,
\label{ohm}
\eeq
where $\sigma$ is the electrical conductivity and $\mbf{E}$ is the electric field 
acting on the plasma at rest. We notice that $\mbf{J'_2} \perp \mbf{\hat{b}}$ while 
$\mbf{J'_1}$ generally includes a field-aligned and a cross-field component. 

Both $\mbf{J_2}$ and $\mbf{J'_2}$ lie fully on the plane $S_{\perp}$ perpendicular 
to the magnetic field vector $\mbf{\hat{b}}$ at any given spatial location. 
Therefore, there is a current density $\mbf{J_{\perp}}$ also lying on $S_{\perp}$ 
such that 
\beq
\mbf{J'_2}=\mbf{J_2}+\mbf{J_{\perp}}\;\;.
\label{gn}
\eeq
Substituting $\mbf{J'_2}$ from equation (\ref{ohm}) into equation (\ref{gn}), one 
finds
\beq
\mbf{u} \times \mbf{\hat{b}}={{\eta} \over {B}} \mbf{j_2} + {{\eta} \over {B}} \mbf{j_{\perp}}\;\;,
\label{gn1}
\eeq
where $\eta = (c^2 / 4 \pi \sigma)$ is the magnetic diffusivity, 
$\mbf{j_2}=(4 \pi /c) \mbf{J_2}$, and $\mbf{j_{\perp}}=(4 \pi /c) \mbf{J_{\perp}}$. 
From equation (\ref{gn1}) one may now infer the horizontal components $u_x$, $u_y$ 
of $\mbf{u}$, namely, 
\begin{eqnarray}
\begin{array}{l}
u_x=(1/b_z)[u_zb_x - (\eta/B)j_{2_y} - (\eta/B)j_{\perp _y}]\\
u_y=(1/b_z)[u_zb_y + (\eta/B)j_{2_x} + (\eta/B)j_{\perp _x}]\;\;,\\
\end{array}
\label{uh}
\end{eqnarray}
where $b_i=(B_i/B)$; $i \equiv \{ x,y,z \}$ are the relative magnetic field strengths 
and the components of $\mbf{\hat{b}}$. An important point regarding equation 
(\ref{gn1}) and its solution, equations (\ref{uh}), is that one of the 
components of the velocity $\mbf{u}$, in the above formulation the vertical velocity 
$u_z$, is left fully {\it unconstrained}. Moreover, all the components of 
$\mbf{j_{\perp}}$, the horizontal components of $\mbf{j_2}$ (see also equation 
(\ref{amp})), and the magnetic diffusivity $\eta$ are unknown. Evidently, therefore, 
$\mbf{u}$ cannot be evaluated by equations (\ref{uh}). Nevertheless, one 
may further decompose $\mbf{u}$ into 
a field-aligned and a cross-field components, $\mbf{u_{||}}$ and $\mbf{u_{\perp}}$, 
respectively. First, we multiply equation (\ref{gn1}) externally by $\mbf{\hat{b}}$ 
to solve for the cross-field velocity, namely, 
\beq
\mbf{u_{\perp}}=-{{\eta} \over {B}} \mbf{j_2} \times \mbf{\hat{b}}  
                -{{\eta} \over {B}} \mbf{j_{\perp}} \times \mbf{\hat{b}}\;\;, 
\label{up1}
\eeq
or, by substituting equation (\ref{amp}) for $\mbf{j_2}$, 
\beq
\mbf{u_{\perp}}={{\eta} \over {B}} (\mbf{\nabla}B)_{\perp} - 
                {{\eta} \over {B}} \mbf{j_{\perp}} \times \mbf{\hat{b}}\;\;,
\label{up}
\eeq
where $(\mbf{\nabla}B)_{\perp}$ is the cross-field component of $\mbf{\nabla}B$.
Now we can use equations (\ref{uh}) and the definition of the field-aligned velocity, 
$\mbf{u_{||}}=(\mbf{u} \cdot \mbf{\hat{b}})\mbf{\hat{b}}$, to obtain 
\beq
\mbf{u_{||}}=[(u_z - u_{\perp _z})/b_z] \mbf{\hat{b}}\;\;,
\label{ufa}
\eeq
where we have used the vertical component of the vector equation (\ref{up1}).

While we cannot directly evaluate $\mbf{u_{||}}$ and $\mbf{u_{\perp}}$ 
from equations 
(\ref{up1}) - (\ref{ufa}), the above expressions are useful because they provide 
clues on a particular prescription for $u_{\perp _z}$ that will be used to enforce 
a unique solution of the ideal induction equation. 

The general equations (\ref{up1}) and (\ref{up}) may give the impression that 
$\mbf{u_{\perp}}=0$ in the ideal MHD limit, where $\eta \rightarrow 0$. This is 
not the case, however: $\mbf{j_{\perp}}$ from equations (\ref{ohm}) and (\ref{gn}) 
scales linearly with the electrical conductivity $\sigma$, i.e., 
$j_{\perp} \propto \sigma \propto (1/\eta)$. Therefore, 
$\lim _{\eta \rightarrow 0} u_{\perp} \propto \lim _{\eta \rightarrow 0} (\eta /\eta)$, 
which does not necessarily vanish. 
\section{Minimum structure velocity field reconstruction}
\subsection{The cross-field MSR velocity}
As noted in \S3, the analytical expressions for the field-aligned and the cross-field 
velocity components, equations (\ref{up1}) - (\ref{ufa}), 
cannot be evaluated in their present form. As 
a first step to evaluate them, let us employ an assumption regarding  
$\mbf{u_{\perp}}$, $\mbf{\nabla}B$, and $\mbf{\hat{b}}$: we assume that the 
cross-field velocity $\mbf{u_{\perp}}$, equation (\ref{up}), 
relates {\it only} to the cross-field magnetic gradients 
$(\nabla B)_{\perp}$ and aims to either sustain or eliminate them. In other words, we 
assume that $\mbf{u_{\perp}}$ lies on the plane defined by $(\mbf{\nabla}B)_{\perp}$ 
and $\mbf{\hat{b}}$ at any given {\it strong-field} 
location. This assumption is not intended for weakly or partially magnetized 
regions and yields the mathematical relation 
\beq
\mbf{\nabla}B \times \mbf{\hat{b}} \cdot \mbf{u}=0\;\;or,\;\;\;equivalently,\;\;\;\;
(\mbf{\nabla}B)_{\perp} \times \mbf{\hat{b}} \cdot \mbf{u_{\perp}}=0\;\;,
\label{cop}
\eeq
since only the cross-field components $(\mbf{\nabla}B)_{\perp}$ and 
$\mbf{u_{\perp}}$ of $\mbf{\nabla}B$ and $\mbf{u}$, respectively, participate 
in the equation. 
Correlating the two cross-field current density components $\mbf{J_2}$ and 
$\mbf{J'_2}$ from equations (\ref{amp}) and (\ref{ohm}), respectively, one finds 
\beq
\mbf{J_2} \times \mbf{J'_2}=-{{\sigma} \over {4 \pi}}
(\mbf{\nabla}B \times \mbf{\hat{b}} \cdot \mbf{u})\mbf{B}\;\;.
\label{ccor}
\eeq
From our assumption in equation (\ref{cop}) one then finds 
$\mbf{J_2} \times \mbf{J'_2}=0$, so $\mbf{J'_2}=k \mbf{J_2}$, where $k$ is a 
real number. From equation (\ref{gn}) one further 
finds $\mbf{j_{\perp}}=(k-1)\mbf{j_2}$ and the cross-field velocity $\mbf{u_{\perp}}$ 
from the general equations (\ref{up1}) or (\ref{up}) becomes 
\beq
\mbf{u_{\perp}}=-{{\eta} \over {B}} k \mbf{j_2} \times \mbf{\hat{b}}\;\;\;\;or\;\;\;\;
\mbf{u_{\perp}}= {{\eta} \over {B}} k (\mbf{\nabla}B)_{\perp}\;\;,
\label{copl}
\eeq
respectively. 

Despite the above simplification furnished by the assumed {\it coplanarity} 
between $\mbf{u_{\perp}}$, $(\nabla B)_{\perp}$, and $\mbf{\hat{b}}$, equations 
(\ref{copl}) still cannot be evaluated. Therefore, we will further employ 
the {\it minimum structure} approximation. 
This approximation has been introduced and discussed extensively 
by Georgoulis, LaBonte, \& Metcalf (2004) and Georgoulis \& LaBonte (2004). The 
idea behind the minimum structure assumption is to minimize the sheath currents 
that are thought to flow peripherally on the surfaces of magnetic flux tubes in the 
low solar atmosphere, thus making the low-lying magnetic fields as 
{\it space-filling} as possible. 
More specifically, the minimum structure approximation minimizes the 
magnitude $J_2$ of the cross-field current density $\mbf{J_2}$ in equation 
(\ref{amp}) by solving for the unknown vertical gradient $(\partial B / \partial z)$ 
of the magnetic field strength. To realize that this is consistent with minimizing 
the surface ``sheath'' currents, consider a magnetic flux tube of finite 
cross-section embedded in a field-free atmosphere. Then, unless $\mbf{\nabla}B$ 
is fully field-aligned, $J_2$ is more intense on the surface of the flux tube, 
where the transition from a magnetized into an unmagnetized medium takes place. 
The magnitude $J_2$ of $\mbf{J_2}$ becomes minimum when 
\beq
{{\partial B} \over {\partial z}} = {{b_z} \over {b_x^2 + b_y^2}} 
( b_x {{\partial B} \over {\partial x}} + b_y {{\partial B} \over {\partial y}})\;\;.
\label{dB_z}
\eeq
Knowledge of $(\partial B / \partial z)$ enables one to calculate the components 
of $\mbf{J_2}$ or $\mbf{j_2}=(4 \pi/c)\mbf{J_2}$ from equation (\ref{amp}). 
Substituting $(\partial B / \partial z)$ from equation (\ref{dB_z}) in the definition 
of $\mbf{J_2}$, equation (\ref{amp}), one finds that the minimum structure 
approximation leads to 
\beq
(\mbf{j_2} \times \mbf{\hat{b}})_z=0\;\;\;\;\;\;or,\;equivalently,\;\;\;\;
(\mbf{\nabla}B)_{\perp _z}=0\;\;.
\label{msr}
\eeq

Substituting equations (\ref{msr}) of the minimum structure approximation into 
equations (\ref{copl}) of the coplanarity assumption one then finds that the 
vertical component $u_{\perp _z}$ of the cross-field velocity becomes zero, i.e., 
\beq
u_{\perp _z}=0\;\;.
\label{uzp}
\eeq

Therefore, a restricted {\it coplanar} minimum structure approximation
results in a purely horizontal cross-field velocity 
$\mbf{u_{\perp}}$. Both the minimum structure approximation and the coplanarity 
assumption are intended for strongly magnetized regions and hence they can be 
combined. Notice, however, that the assumption $u_{\perp _z}=0$ is principally 
{\it inconsistent} with 
magnetic flux emergence or submergence, because these effects take place along 
polarity reversal lines, where $B_z=0$, so $u_{\perp _z}$ has to be nonzero 
by definition in these locations. We believe that 
our approximation can be justified, however, if one considers the length and 
time scales involved in flux emergence or submergence. These issues are 
discussed in detail in \S6.1, but in summary the 
effects of flux emergence/submergence 
generally require length scales much smaller and time 
scales much larger than the respective scales of interest in 
an inductive velocity field calculation. Nonetheless, the coplanar minimum structure 
approximation is expected to perform poorly in small-scale magnetic features 
characterized by significant flux emergence or submergence. 

One might think that the coplanarity assumption, equation (\ref{cop}), provides an 
additional constraint for $u_{\perp _z}$, besides equation (\ref{uzp}). This would 
invalidate our coplanar minimum structure approximation since the system of equations 
would then become overdetermined (equations (\ref{iz}), (\ref{uzp}), 
$\mbf{u_{\perp}} \cdot \mbf{B}=0$, and equation (\ref{cop}) for the three 
components of $\mbf{u_{\perp}}$). This is not the case, however: the coplanarity 
assumption, equation (\ref{cop}), combined with the orthogonality condition, 
$\mbf{u_{\perp}} \cdot \mbf{B}=0$, can only give $u_{\perp _z} =0$ in the minimum 
structure approximation. To realize this, one may write equation (\ref{cop}) as 
follows:
\beq
u_{\perp _z}=-{{1} \over {(\nabla B \times {\mbf B})_z}} 
[u_{\perp _x} (\nabla B \times {\mbf B} )_x + 
 u_{\perp _y} (\nabla B \times {\mbf B} )_y]\;\;.
\label{R1}
\eeq
Substituting equation (\ref{dB_z}) of the minimum structure assumption into 
equation (\ref{R1}) one finds 
\beq
u_{\perp _z}={{B_z} \over {B_h^2}} {\mbf B_h} \cdot {\mbf u_{\perp _h}}\;\;.
\label{R2}
\eeq
Now, from the orthogonality condition between $\mbf{u_{\perp}}$ and $\mbf{B}$ 
one obtains 
\beq
u_{\perp _z}=-{{1} \over {B_z}} {\mbf B_h} \cdot {\mbf u_{\perp _h}}\;\;.
\label{R3}
\eeq
Combining equations (\ref{R2}) and (\ref{R3}) 
for nonzero $B_z$ and $\mbf{B_h}$, one finds 
\beq
B_z^2 {\mbf B_h} \cdot {\mbf u_{\perp _h}} = - B_h^2 
{\mbf B_h} \cdot {\mbf u_{\perp _h}}\;\;, 
\label{R4}
\eeq
which is only true if ${\mbf B_h} \cdot {\mbf u_{\perp _h}}=0$. The only case 
where this is valid is when $u_{\perp _z}=0$, as readily shown 
from equations (\ref{R2}) and (\ref{R3}). Therefore, the coplanar minimum 
structure assumption closes the system of equations for $\mbf{u_{\perp}}$ 
which can now be solved exactly and uniquely. 
\subsection{The field-aligned MSR velocity}
Using our special case of a coplanar minimum structure approximation, 
$u_{\perp _z}=0$, the vertical velocity $u_z$ from equation (\ref{uz}) simplifies to 
\beq
u_z={{B_z} \over {B_l}} (u_l - \alpha u_{\perp _x} - \beta u_{\perp _y})\;\;. 
\label{uz_msr}
\eeq
Equation (\ref{uz_msr}) will be used in the following 
to calculate the field-aligned velocity $\mbf{u_{||}}$
in the coplanar minimum structure approximation 
by means of equation (\ref{ufa}).
\section{Reconstruction of the MSR velocity field in solar active regions}
To calculate the MSR velocity field vector in solar active regions we will use vector 
magnetogram data obtained by the Imaging Vector Magnetograph 
(IVM; Mickey et al. 1996) of the University of 
Hawaii's Mees Solar Observatory. 
The IVM records the complete Stokes vector at each of 30 
spectral points through the Fe {\small I} $6302.5$ \AA$\;$photospheric spectral line. 
The magnetic field components are 
obtained via an inversion code that includes LTE radiative transfer, magneto-optic 
effects, and the filling factor of the unresolved flux tubes 
(Landolfi \& degl'Innocenti 1982). 

The above inversion code also determines the Doppler shift of the spectral line, and 
hence the longitudinal Doppler velocity. The fit to the Stokes I profile determines 
the line-center wavelength of the Fe {\small I} line with velocity $V_I$. The fit 
to the Stokes Q, U, and V profiles for the magnetic field components determines the 
velocity $V_m$ of the magnetized areas and provides the wavelength difference from the 
Stokes I solution with velocity $\Delta V_m=V_m - V_I$. The longitudinal velocity 
$u_l$ is then calculated by the relation 
\beq
u_l=V_I + \Delta V_m\;\;.
\label{IVM_Dop}
\eeq
The instrumental spatial quadratic variation of wavelength, caused by the off-axis 
passage through the Fabry-Perot etalon, is removed at the end. Equation (\ref{IVM_Dop}) 
provides the correct Doppler velocity for magnetized areas and accounts for the 
potential error described by Chae, Moon, \& Pevtsov (2004) where using the shift 
$V_I$ of Stokes I at line center risks mixing the HD longitudinal motions of the 
unmagnetized plasma with the MHD motions of the magnetized plasma. Nevertheless, 
there is still space for improvement for the IVM Doppler velocities 
because an error in the reduction of the magnetograms tends to weigh 
$\Delta V_m$ by the filling factor $f$. As a result, 
$\Delta V_m$ in equation (\ref{IVM_Dop}) corresponds more to $f \Delta V_m$. The effect 
is minimal in our calculations since we focus on strong-field regions such 
as sunspots and strong-field plages, where $f \simeq 1$, but it needs to be addressed 
in any case. In addition, we average the 
inferred Stokes images for timescales of $20\;min$ to $30\;min$ to eliminate 
oscillatory effects, such as the $5-min$ oscillations, from the inferred Doppler 
velocity $u_l$. 

The $180^o$-ambiguity for each magnetogram of a given pair has been resolved using the 
structure minimization technique of Georgoulis, LaBonte, \& Metcalf (2004). After the 
ambiguity is resolved the magnetograms are co-aligned with respect to their vertical 
components. Besides eliminating oscillatory effects in the Doppler velocity, averaging 
over periods of $20-30\;min$ enhances the signal-to-noise ratio in both the inferred 
magnetic field components and the longitudinal velocity. 
We use noise thresholds of $100\;G$ and $200\;G$ 
for the averaged vertical and horizontal magnetic fields, respectively. 
The calculations described in \S\S 2 and 4 
are performed using the heliographic magnetic field components on the image plane, 
but the results can be readily transferred to the local, heliographic, plane. In the 
following, we discuss three examples of active regions (ARs) that 
showed distinctive flows. 
\subsection{NOAA AR 9114}
A series of vector 
magnetograms of NOAA AR 9114 were recorded by the IVM on 2000 August 8. 
Figure \ref{f1} shows the average ambiguity-free 
magnetic field vector (Figure \ref{f1}a) and Doppler 
velocity (Figure \ref{f1}b) for a pair of vector magnetograms obtained at 19:31 UT and 
at 19:59 UT. 

In Figure \ref{f2} we illustrate the convergence process in calculating 
electrostatic field $\mbf{G}$. As discussed in 
\S2.3, the use of Fourier transforms does not provide results with accuracy up to 
the machine accuracy, unlike the ADI or SOR techniques. 
In Figure \ref{f2} the error in 
minimizing $\mathcal{R}$ ($\mathcal{R} \simeq 0$) is $\sim 7$\% because 
$\mbf{B}$ is somewhat noisy, despite the averaging. 

Given the approximate calculation of the cross-field velocity $\mbf{u_{\perp}}$, 
it is interesting to compare the observed temporal variation 
$(\Delta B_z / \Delta t)$ of the vertical magnetic field $B_z$ between the two 
magnetograms with the reproduced temporal variation using the solution 
for $\mbf{u_{\perp}}$. 
This can be done by advancing the average $B_z$ in time for a time 
difference $(\Delta t /2)$,  
using the velocity field $\mbf{u_{\perp}}$ in the induction equation (\ref{iz}). 
Unfortunately, we cannot do the same for the horizontal components $B_x$, $B_y$ of 
$\mbf{B}$, since this requires knowledge of the unknown height derivatives 
$(\partial B_x / \partial z)$, $(\partial B_y / \partial z)$ besides $\mbf{u_{\perp}}$. 
If these derivatives were known, one would 
then be able to advance the entire magnetogram from time 
$t+(\Delta t /2)$ to any reasonable time interval $\Delta t'$ 
and thus predict the temporal evolution of 
the magnetic field vector in compliance with the induction equation. The 
observed and the reproduced temporal variations $(\Delta B_z / \Delta t)$ are shown in 
Figures \ref{f3}a and \ref{f3}b, respectively. Notice that the observed 
$(\Delta B_z / \Delta t)$ is reproduced in remarkable detail, although a careful 
inspection will 
reveal that the observed variation is noisier than the reproduced variation. This 
should be expected from the use of Fourier transforms. 
A correlation between the observed and the reproduced 
temporal variation is shown in Figure \ref{f3}c. 
The correlation coefficient between the two quantities 
is $\sim 0.95$ for both the linear and the non-parametric 
(Spearman) rank estimations. The best linear fit to the scatter plot, 
indicated by the dashed line, shows a slope $1.08 \pm 0.01$ 
and it is very similar to the theoretical relation 
$(\Delta B_z / \Delta t) _{observed}=(\Delta B_z / \Delta t) _{reproduced}$, shown 
by the solid line. 

The error in the convergence process for $\mbf{G}$ and hence in the 
final $\mbf{u_{\perp}}$ is not large enough to significantly impact the results. 
For a moderate $(\Delta B_z / \Delta t)_{observed}=0.1\;G\;s^{-1}$ the best linear 
fit of Figure \ref{f3}c shows an expected 
$(\Delta B_z / \Delta t)_{reproduced} \simeq 0.092\;G\;s^{-1}$, 
while for a large $(\Delta B_z / \Delta t)_{observed}=0.2\;G\;s^{-1}$ 
the expected reproduced value is $\sim 0.184\;G\;s^{-1}$. Therefore, the expected 
mean fractional error 
$(|(\Delta B_z / \Delta t)_{observed} - (\Delta B_z / \Delta t)_{reproduced}|/
|(\Delta B_z / \Delta t)_{observed}|)$ is $\sim 8$\%, similar to the error in finding 
$\mathcal{R} \simeq 0$. The reproduced values are systematically slightly smaller 
than the observed values. Large observed 
$(\Delta B_z / \Delta t)$ may occur because of noise or errors in the magnetic field 
measurements, subtle inaccuracies in the co-alignment of the two magnetograms or 
changes in the seeing conditions in the course of the ground-based observations and 
hence they cannot be reproduced using Fourier transforms. Further testing 
the accuracy of the $\mbf{u_\perp}$-solution we find that it is consistent with the 
coplanarity assumption, equation (\ref{copl}), with a mean fractional error 
$[|\mbf{u_\perp} \cdot (\nabla B \times \mbf{B})|/(u_{\perp} |\nabla B \times \mbf{B}|)]$
of $\sim 6.2$\%. 
   
NOAA AR 9114 was chosen for this study because of the distinctive flows observed around 
the AR's leading sunspot with positive magnetic polarity. This sunspot 
exhibited both a counterclockwise rotation during the observing interval 
(see also Brown et al. 2003) and 
intense flux-dispersing sunspot outflows characterized by the appearance and 
disappearance of numerous short-lived moving magnetic features (see Harvey \& 
Harvey 1973; Nindos \& Zirin 1998 for classical descriptions). 
It is interesting to check whether the rotation and outflows are reproduced 
by the MSR velocity solution. The solution is 
shown in Figure \ref{f4} with the horizontal velocity vector plotted on top of the 
average vertical magnetic field (Figure \ref{f4}a) and the 
vertical MSR velocity (Figure \ref{f4}b), i.e., the velocity $u_z=u_{|| _z}$, 
since $u_{\perp _z}=0$. Only areas of strong magnetic 
field are shown in the images. 
The dashed box in Figure \ref{f4}a indicates the sub-region occupied 
by the sunspot. As we can see from Figure \ref{f4}, the 
sunspot's rotation and radial outflows are 
reproduced nicely. Outflows occur with a velocity ranging between $0.4\;km\;s^{-1}$ and 
$1\;km\;s^{-1}$. The calculated vertical velocity reveals mostly upflows 
reaching up to $0.5\;km\;s^{-1}$.

In Figure \ref{f5} we focus on the sunspot of the AR. The MSR velocity field has been 
calculated for three different magnetogram pairs of the AR over a period of 
$\sim 4\;hr$. We show the cross-field velocity $\mbf{u_{\perp}}$ (upper row of 
images) and the total velocity $\mbf{u}$ (lower row of images). In all images, the 
grayscale background is the vertical velocity $u_z$. 

The counterclockwise rotation of the sunspot observed by Brown et al. (2003) 
and inferred from the IVM magnetogram movie is evidently reproduced by the 
cross-field velocity (Figures \ref{f5}a to \ref{f5}c). The 
rotation involves the outer penumbra of the sunspot, while the inner penumbra 
and/or the outer edge of the umbra show a weaker opposite (clockwise) rotation. 
This, opposite to the prevailing, sunspot rotation is more evident in Figure \ref{f5}c. 
Moving progressively from Figure \ref{f5}a to Figure \ref{f5}c we notice that 
the counterclockwise rotation of the sunspot appears to subside with time. 

While the cross-field velocity is responsible for the sunspot rotation in 
NOAA AR 9114, the sunspot outflows are contributed by the field-aligned velocity 
(Figures \ref{f5}d to \ref{f5}f). The outflows persist at the northern part of the 
sunspot, but they appear to subside at the southern 
part of the sunspot. It is not easy to visually determine from the 
magnetogram movie 
whether the outflows at the southern part weaken during the end of IVM 
daily observing interval, while 
we can see them clearly at the northern part, 
as the northward flows are more intense and 
systematic than the southward ones. In addition, we noticed that the seeing conditions 
deteriorate toward the end of the observing interval, so weaker flows are 
probably harder to pick up. From Figures \ref{f5}e, \ref{f5}f 
a change is also evident in the vertical component 
of the MSR velocity, as compared to Figure \ref{f5}d. Upflows of the order 
$0.5\;km\;s^{-1}$ characterize the umbra and part of the penumbra until $\sim$18:00 UT. 
These upflows have decreased significantly at $\sim$20:00 UT and they continue to 
decrease at $\sim$22:00 UT. 

In general, we find that the most distinctive flow features in NOAA AR 9114, 
namely the sunspot rotation and outflows, are 
believably reproduced by the MSR velocity field solution. Moreover, Figure \ref{f5} 
shows the consistency of our velocity solution between different pairs of vector 
magnetograms, given the uncertainties and a likely change in the seeing conditions 
toward the end of the IVM observing interval. 
The inferred sunspot rotation indicates that left-handed twist is injected 
in the active-region atmosphere via the emergence of a plasma-carrying helical 
magnetic structure. 
Smaller, concentric, toroidal structures with opposite chirality may also 
emerge simultaneously, as indicated by the weaker opposite rotation 
seen in the inner 
parts of the penumbra or in the outer umbra. 
On the other hand, the Evershed outflows of the upwelled plasma occur as a 
result of the dynamic pressure gradients built up during 
the emergence process. Because this 
plasma is mostly magnetized, outflows occur mainly along the emerging magnetic field 
lines so they are captured by the field-aligned velocity solution. Both the sunspot 
rotation and the outflows appear to weaken in the course of time, at least for 
the southern part of the sunspot. If this is not an artifact owning to the 
variable seeing conditions, a possible clue might be provided 
by the decreasing vertical upflow velocity, 
which indicates that the emergence of the helical magnetic structure and the 
upwelling of the plasma gradually subside with time.  
\subsection{NOAA AR 8210}
The passage of NOAA AR 8210 from the 
visible solar disk early in May 1998 resulted in a series of 
solar flares and at least one halo CME on 1998 May 2 (Warmuth et al. 2000; 
Pohjolainen et al. 2001; Sterling \& Moore 2001). 
The AR is a well-studied subject from 
photosphere to corona (Thompson et al. 2000; Wang et al. 2002; 
Roussev et al. 2004) and a case subject for the Solar MURI Project 
(Fisher et al. 2003) because of its dynamical activity linked to a 
distinctive $\delta$-sunspot photospheric magnetic 
configuration. Intense photospheric flows were also observed. As a result, 
NOAA AR 8210 was the subject of the velocity field calculation in 
the studies of Welsch et al. (2004) and Longcope (2004). 

The IVM obtained a series of vector magnetograms of the AR on 1998 May 1. For a pair of 
these magnetograms, taken at 17:20 UT and at 17:52 UT, 
the average ambiguity-free magnetic 
field vector and the average Doppler velocity are given in Figures \ref{f6}a and 
\ref{f6}b, respectively. A complex $\delta$-sunspot configuration in the AR 
is evident from Figure \ref{f6}a. Inspecting the IVM magnetogram movie we notice 
counterclockwise motions in the strong positive-polarity field adjacent to the 
eastern part of the sunspot. The spot itself appeared to move clockwise along the 
polarity inversion line, 
which presumably helped accumulating a significant magnetic shear. 
The western and southwestern parts of the sunspot exhibited significant outflows, while 
a mixed picture of inflows and outflows occurred at the southeastern 
part of the sunspot. Moreover, there was some 
flux emergence northwest of the sunspot with a small negative-polarity 
pore emerging and moving rapidly to the southwest. 

We calculated the MSR velocity for the 
above pair of magnetograms. In this example, 
the convergence to the unique minimum structure solution (i.e. $\mathcal{R} \simeq 0$; 
Figure \ref{f2} for the previous example) was achieved with an error 
of $\sim 4.7$\%. The observed and the reproduced temporal variation 
$(\Delta B_z / \Delta t)$ are given in Figures \ref{f7}a and \ref{f7}b. Again we note 
that the observed variation is reproduced closely, although it 
is somewhat noisier than the reproduced variation. 
Correlating the observed and reconstructed 
temporal variations (Fig. \ref{f7}c) 
we calculate a high correlation coefficient ranging 
between $0.97$ and $0.98$, while the best linear fit of the scatter plot 
(dashed line; slope $1.05 \pm 0.005$) is 
very similar to the theoretical relation of equality between the two temporal variations 
(solid line). From the best fit we find that for a commonly observed vertical field 
variation of $0.1\;G\;s^{-1}$  the expected reproduced variation is 
$\sim 0.95\;G\;s^{-1}$, 
while for a large observed variation of $0.2\;G\;s^{-1}$ the expected reproduced 
variation is $\sim 0.19\;G\;s^{-1}$. As with NOAA AR 9114, the expected mean fractional 
error $(|(\Delta B_z / \Delta t)_{observed} - (\Delta B_z / \Delta t)_{reproduced}|/
|(\Delta B_z / \Delta t)_{observed}|)$ is $\sim 5$\%, similar to the error in 
$\mathcal{R} \simeq 0$. Moreover, the mean fractional error 
$[|\mbf{u_\perp} \cdot (\nabla B \times \mbf{B})|/(u_{\perp} |\nabla B \times \mbf{B}|)]$
in the fulfillment of the coplanarity assumption is $\sim 5$\%. 

The MSR velocity field solution for the $\delta$-sunspot is shown in Figure \ref{f8}. 
We show both the cross-field velocity solution $\mbf{u_{\perp}}$ 
(Figures \ref{f8}a, \ref{f8}c) and the total calculated velocity $\mbf{u}$ 
(Figures \ref{f8}b, \ref{f8}d). We notice that both the sunspot outflows and part 
of the counterclockwise motion, especially in the northern part of the 
positive-polarity arm east of the 
$\delta$-sunspot, are nicely shown. Sunspot outflows occur with a velocity 
$\sim (0.5 - 0.8)\;km\;s^{-1}$, while the positive flux accumulations at the east 
move with a velocity $\sim (0.2 - 0.5)\;km\;s^{-1}$. 
Some velocity shear along the polarity inversion line can be also seen. 
Calculations on, or very close to, the line are not possible because 
$|B_z| \rightarrow 0$, so we can only calculate the flows at the 
vicinity of the line. We also notice some inflows with velocity 
$\sim (0.5 - 0.8)\;km\;s^{-1}$ at the southeastern part of the $\delta$-sunspot. 
The MSR solution shows these motions as both outflows and inflows for different 
pairs of magnetograms. This is probably because the fine mixture of inflows and 
outflows inferred from the IVM magnetogram movie in this area may not 
be reproduced reliably by the MSR solution. 
The clockwise rotation of the $\delta$-sunspot, also inferred 
from the $H \alpha$ observations of Warmuth et al. (2000), 
is not evident in the MSR velocity solution. 

From a visual comparison of the MSR solution, the ILCT solution of Welsch et al. 
(2004), and the MEF solution of Longcope (2004) we find that the strong penumbral 
inflows and outflows seen in the IVM magnetogram movie are reproduced by the MSR 
velocity but they are not a conspicuous feature of the ILCT and MEF solutions. 
The counter-clockwise motions of the positive-polarity arm of the sunspot, however, 
appear more or less in all three solutions. In terms of the calculated flow amplitude, 
the MSR solution leads to larger horizontal flows than both the ILCT and the MEF 
solutions. The strongest flows for these solutions appear to be $\sim 0.4\;km\;s^{-1}$ 
and $\sim 0.8\;km\;s^{-1}$, respectively, while our calculated flows can be as fast 
as $\sim (1 - 1.5)\;km\;s^{-1}$. It is not clear whether the MSR approach 
overestimates or the ILCT and the MEF approaches underestimate the horizontal 
cross-field velocity. Assuming $u_{\perp _z}=0$ in the MSR technique may introduce 
additional horizontal flows to solve the induction equation. 
On the other hand, the ILCT solution relies on an 
essentially smoothed LCT velocity and the MEF solution explicitly demands minimal 
flows, so these two solutions may underestimate the 
horizontal flows. Regarding the small emerging flux region 
at the northwest of the $\delta$-sunspot, indicated by the white oval in Figure 
\ref{f8}a, only the ILCT solution reproduces the 
intense flows to the southwest. The MEF solution does not include this part of the 
AR in the analysis, while the MSR solution cannot reproduce these flows. 
There are two reasons for this discrepancy: first, 
as explained in the Introduction, any inductive technique 
assisted by tracking is more likely to reproduce such flows which may in part be 
apparent flows, caused by the emergence of inclined magnetic structures. Second, the 
main assumption behind the MSR solution, $u_{\perp _z}=0$, may break down locally in 
emerging flux regions as mentioned in \S4.1 and discussed in detail in \S6.1. 

In Figure \ref{f9} we show the MSR velocity in the $\delta$-sunspot for three 
different magnetogram pairs of the AR. The upper row of images  
(Figures \ref{f9}a - \ref{f9}c) show the MSR horizontal velocity plotted on top of the 
vertical magnetic field. In the lower 
row of images the grayscale background is the vertical component of the MSR velocity. 
The three magnetogram pairs cover a period of $\sim 2.5\;hr$. 
The MSR velocity solutions are similar in all three cases: 
both the counterclockwise motion of the positive polarity arm at the 
east of the sunspot and the penumbral inflows/outflows are similarly reproduced in all 
images while the solution for the vertical velocity changes only slightly. 
\subsection{NOAA AR 10030}
NOAA AR 10030 was also a flaring AR involving large amounts of 
magnetic flux in a complex 
multipolar magnetic configuration. The most important event associated with this AR 
was a X3 flare followed by a halo CME late on 2002 July 15. 
We have calculated the MSR velocity field in the AR for a pair of vector 
magnetograms recorded by the IVM at 19:50 UT and at 20:21 UT on 2002 July 15. Figure 
\ref{f10}a depicts the average, ambiguity-free, vertical magnetic field for the 
magnetogram pair, while Figure \ref{f10}b shows the average Doppler velocity. 
Inspecting the IVM magnetogram movie of the AR late on 2002 July 15, we notice
(i) a possible clockwise rotation of the westernmost leading sunspot of the 
AR (ovals A1 and B1 in Figures \ref{f10}a and \ref{f10}b, respectively) and 
(ii) a highly sheared neutral line close to the trailing spot of the AR at the east 
(ovals A2 and B2 in Figures \ref{f10}a and \ref{f10}b, respectively). The sheared 
neutral line is also associated with 
an intense upflow/downflow pattern in the longitudinal 
Doppler velocity with flow amplitudes $(0.6-0.8)\;km\;s^{-1}$. 

In Figure \ref{f11} we show the observed (Figure \ref{f11}a) and the reproduced 
(Figure \ref{f11}b) temporal variations of the vertical magnetic field for the 
magnetogram pair. The observed temporal variation is somewhat noisier 
than the reproduced one, as is the case in the previous examples. In this particular 
example the convergence to the MSR solution was achieved with an error of 
$\sim 6.1$\% for $\mathcal{R} \simeq 0$. 
The observed and the reproduced temporal 
variations are correlated in the scatter plot of Figure \ref{f11}c. Both the linear and 
the Spearman rank correlation coefficient are $\sim 0.94$ and the best linear fit 
(dashed line; slope $1.06 \pm 0.008$) 
is close to the theoretical expression of equality between the two temporal 
variations (solid line). The mean fractional error between 
$(\Delta B_z / \Delta t)_{observed}$ and $(\Delta B_z / \Delta t)_{reproduced}$ 
is $\sim 6$\%, while the mean fractional error in the fulfillment of the 
coplanarity assumption is $\sim 4.4$\%. 

In Figure \ref{f12} we depict the MSR velocity solution for the AR. The horizontal 
velocity is plotted on top of $B_z$ (Figure \ref{f12}a) and on top of $u_z$ 
(Figure \ref{f12}b). Given the significant complexity of the magnetic configuration, 
individual features of the flow are hardly discernible. For this reason we focus on 
the two most conspicuous 
flow features of the IVM magnetogram movie, namely the clockwise 
rotation of the westernmost leading sunspot and the sheared 
neutral line at the east. 
The two corresponding sub-areas of the MSR velocity solution of Figure \ref{f12} are 
magnified and shown in details (A1), (B1) (sunspot) and (A2), (B2) (neutral line) 
in Figure \ref{f12}. The clockwise rotation of the sunspot is evidently verified by 
the MSR velocity solution: the sunspot rotates with a velocity 
$\sim(0.3\;-\;0.6)\;km\;s^{-1}$. On the other hand, the MSR velocity solution 
reproduces part of the velocity shear at the vicinity of the neutral line
but it does not show a fully sheared neutral line as one concludes from the 
magnetogram movie. As already mentioned, calculations on, or very close to, the 
neutral line are not possible because $|B_z| \rightarrow 0$. Shearing motions occur 
with a velocity $\sim(0.4\;-\;0.9)\;km\;s^{-1}$. In accordance with the Doppler 
velocity, the MSR vertical velocity also shows intense upflows/downflows with an 
amplitude of $\sim (0.7\;-\;0.8)\;km\;s^{-1}$. Summarizing, the MSR velocity 
reproduces reasonably the most characteristic flows in the AR.  

The inductive velocity can give rise to additional information besides being 
utilized for 3D MHD simulations or for helicity flux calculations. 
For instance, one may partially calculate the 
{\it vorticity} $\mbf{\zeta} = \mbf{\nabla} \times \mbf{u}$ associated with 
the flow. Just like the electric current density, only the vertical component 
of the vorticity, i.e., 
$\zeta _z = [(\partial u_y / \partial x) - (\partial u_x / \partial y)]$, 
can be calculated for single-height magnetic field observations and the respective 
velocity field vector. Even this limited vorticity information, however, can 
provide some knowledge of the temporal evolution on the plane of the calculations. 
A consistent sign of vorticity over a magnetic feature indicates vortex motions 
and whirling flows, while different signs of vorticity indicate velocity shear. 
There are strong indications that vortex motions are present in the 
convection zone on in shallow sub-photospheric layers (Kosovichev 2002; 
L\'{o}pez Fuentes et al. 2003), while it has been shown that Parker's twisting and 
braiding of the magnetic field lines above the photosphere can be achieved by 
small-scale photospheric vortex motions overshooting into the corona 
(Meytlis \& Strauss 1993; Lionello et al. 1998). The 
vertical vorticity for NOAA AR 10030 is given in Figure \ref{f13}. The leading 
sunspot in the west (oval (a)) is mostly characterized by a unique sign of vorticity 
($\zeta _z < 0$) which indicates a clockwise rotation, 
as also shown in Figure \ref{f12}.  
For the other sub-regions of the AR, indicated by ovals (b), (c), and (d), there is 
evidence for both signs of vorticity, so velocity shear is at work in these areas. 
The sheared neutral line discussed above is one of them, indicated by oval (d). 
It is well known that the magnetic complexity increases in sheared regions and 
the likelihood of a flare-triggering instability is high. Indeed, the origin location 
of the X3 flare that was triggered in the AR is an area of strong vortex motions and 
possible shear, indicated by oval (b) in Figure \ref{f13} 
(see Liu et al. (2003) and Gary \& Moore (2004) for more detailed discussions).

A complete description of the evolution in NOAA AR 10030 using the MSR velocity 
lies beyond the scope of this study. From this example, 
however, we illustrate both the importance of knowing the velocity field together 
with the magnetic field vectors in a complex AR 
and that the MSR velocity field solution 
is detailed enough to be used in further analyses. 
\section{Discussion and Conclusions}
Several important tasks in solar physics, from the 
anelastic (non-force-free) 3D MHD modeling 
of the magnetic fields in the corona to the calculation of the magnetic helicity flux 
or the quantitative assessment of flow patterns and vortex motions in the solar 
photosphere/chromosphere 
require simultaneous, reliable, velocity field information besides the 
measured magnetic field vector. The measured LCT velocity alone is 
insufficient for the study of flows in solar active regions 
and bears little agreement with the time-dependent MHD equations. An almost unanimous 
consensus calls for a velocity field calculation in compliance with the MHD theory 
and, in particular, with the induction equation. To calculate an inductive velocity, 
however, one needs reliable {\it vector} magnetograms, contrary to the LCT velocity 
that requires either longitudinal magnetograms or continuum images. This fact 
restricts the applicability of the induction equation. 
In addition, single-height magnetic field measurements 
allow only a partial use of the induction equation which makes the calculation 
of the cross-field velocity an under-determined problem. Additional hypotheses are, 
therefore, required to enforce a unique solution of the ideal induction equation. Three 
previous inductive 
techniques (Kusano et al. 2002; Welsch et al. 2004; Longcope 2004) have so 
far addressed the problem, each relying on its own assumptions. None of the above 
techniques attempts to calculate the field-aligned velocity besides the 
cross-field velocity. 

In this study we attempt to generalize the problem (i) by introducing a general 
solution formalism 
for the vertical component of the ideal induction equation, and (ii) by 
introducing a method to calculate the field-aligned flows 
using the available Doppler velocity information together with the solution of the 
induction equation. The first action allows the use of any additional constraint 
to uniquely solve the ideal induction equation. 
Therefore, a multitude of inductive velocity solutions can be realized by means of 
our general formalism. 
The second action allows a complete 
reconstruction of the velocity field vector that includes the field-aligned velocity, 
provided that the Doppler velocity has been processed adequately to eliminate known 
systematic errors and biases. To show the feasibility of our general approach we 
calculate a unique inductive velocity relying on our own additional constraint, 
which is a coplanar minimum structure approximation. 
We avoid using the LCT velocity and 
we do not explicitly demand minimal flows, 
although we do not consider these assumptions unreasonable by any means. 
In \S6.1 we discuss 
the physics of the coplanar minimum structure approximation and in \S6.2 we 
summarize our results for the MSR velocity in the three studied solar active 
regions.  
\subsection{Feasibility of the coplanar minimum structure approximation}
As discussed in \S4.1 the MSR cross-field velocity $\mbf{u_{\perp}}$ (i) 
relates only to the cross-field magnetic 
gradients $(\nabla B)_{\perp}$ (equation (\ref{copl})), 
and (ii) lies on the horizontal 
plane. There is no additional constraint in the relation of $\mbf{u_{\perp}}$ 
and $(\nabla B)_{\perp}$ so the cross-field velocity can either 
sustain $(k > 0)$ or relax ($k <0$) 
the cross-field magnetic gradients. Illustrative examples of both actions of the 
velocity field can be found (i) in supergranules observed within active regions, 
where the supergranular boundaries are sustained by persistent converging flows 
caused by flux emergence at the center of neighboring cells 
(see, e.g., Georgoulis et al. 2002; Bernasconi et al. 2002), and (ii) in sunspots, 
where strong inflows/outflows are seen in several cases 
(e.g. our results for NOAA ARs 9114 and 8210). The sunspot 
outflows are caused by pressure gradients and 
tend to disperse strong magnetic field gradients at the 
penumbral edges caused by the magnetic flux accumulation and the upwelling of 
magnetized plasma. 

To determine whether our coplanar minimum structure approach is compatible with 
fundamental laws we consider the MHD momentum equation, where 
the Lorentz force has been decomposed into a magnetic pressure 
force $(1/(8 \pi)) \mbf{\nabla}B^2 = (1/(4 \pi))B \nabla B$ and a magnetic tension 
force $(1/(4 \pi)) (\mbf{B} \cdot \mbf{\nabla})\mbf{B}$ (e.g., Jackson 1962), i.e. 
\beq
{{\partial \mbf{u}} \over {\partial t}} + (\mbf{u} \cdot \nabla) \mbf{u} = 
- \nabla P + {{1} \over {4 \pi}} (\mbf{B} \cdot \nabla) \mbf{B} - 
{{B} \over {4 \pi}} \nabla B + \rho \mbf{g}\;\;,
\label{meq}
\eeq
where $P$ is the 
gas pressure, $\rho$ is the mass density, and $\mbf{g}=-g \mbf{\hat{z}}$ is the 
gravitational acceleration. Viscous forces are ignored in equation (\ref{meq}) 
in line with our ideal MHD approach. 
We seek to determine whether the MSR flows 
that result from the coplanar minimum structure approximation 
can balance the magnetic and non-magnetic forces of equation (\ref{meq}) in any given 
case. The analysis is simplified if we introduce a moving orthonormal coordinate 
system with the local axes defined by $\mbf{\hat{b}}$,  
$\mbf{\hat{b}_1}=[(\nabla B)_{\perp}/|(\nabla B)_{\perp}|]$, and 
$\mbf{\hat{b}_2}=\mbf{\hat{b}} \times \mbf{\hat{b}_1}$. 
Then the magnetic field vector, the magnetic gradients, and the velocity field vector 
in the coplanar minimum structure approximation can be written as 
\beq
\mbf{B}=B \mbf{\hat{b}}\;\;,\;\;\;\;
\nabla B = (\nabla B \cdot \mbf{\hat{b}}) \mbf{\hat{b}} + 
(\nabla B)_{\perp} \mbf{\hat{b}_1}
\;\;,\;\;\;\;and\;\;\;\;
\mbf{u}=(\mbf{u} \cdot \mbf{\hat{b}}) \mbf{\hat{b}} + 
s(u_{\perp}) u_{\perp} \mbf{\hat{b}_1}\;\;,
\label{set}
\eeq
respectively, where $s(u_{\perp})= k/|k|$ (equation (\ref{copl})). 
Now the MHD momentum equation can be decomposed in its three 
components along $\mbf{\hat{b}}$, $\mbf{\hat{b}_1}$, and $\mbf{\hat{b}_2}$. 
It can be readily shown that MSR flows can be 
generated or modified ($(\partial \mbf{u} / \partial t) \ne 0$)
by magnetic and non-magnetic 
forces acting along $\mbf{\hat{b}}$ and $\mbf{\hat{b}_1}$ since all terms in 
equation (\ref{meq}) have, in principle, nonzero components along 
$\mbf{\hat{b}}$ and $\mbf{\hat{b}_1}$. Along $\mbf{\hat{b}_2}$, however, the MSR 
flows do not have a component, so $(\partial \mbf{u} / \partial t)=0$.  
In this case, equation (\ref{meq}) becomes 
\beq
(\mbf{u} \cdot \nabla) \mbf{u} \cdot \mbf{\hat{b}_2}= - 
{{\partial P} \over {\partial b_2}} + 
{{1} \over {4 \pi}} (\mbf{B} \cdot \nabla) \mbf{B} \cdot \mbf{\hat{b}_2} 
-\rho g \mbf{\hat{z}} \cdot \mbf{\hat{b}_2}\;\;.
\label{b2}
\eeq
Although $\mbf{u}$, $\nabla B$, and $\mbf{B}$ do not have a component along 
$\mbf{\hat{b}_2}$ (equations (\ref{set})), the dynamic flow pressure 
$(\mbf{u} \cdot \nabla) \mbf{u}$ and the magnetic tension 
$(\mbf{B} \cdot \nabla) \mbf{B}$ have nonzero 
components along $\mbf{\hat{b}_2}$ due to curvature effects. 
As a result, the MSR flows give rise to pressure forces along the 
$\mbf{\hat{b}_2}$-direction but a velocity generation or change 
$(\partial \mbf{u} / \partial t)$ along $\mbf{\hat{b}_2}$ due to magnetic and 
non-magnetic forces is not allowed because the coplanar minimum structure 
approximation precludes any cross-field flows $\mbf{u_{\perp}}$ along $\mbf{\hat{b}_2}$.
This is a limitation of the approximation, so the MSR flows 
are expected to suppress some of the actual inductive flows and hence generate some 
artificial (horizontal, by construction [$u_{\perp _z}=0$]) cross-field 
flows to satisfy the ideal induction equation. The extent of the approximation  
$u_{\perp _z}=0$ determines the magnitude of the artificially introduced horizontal 
flows. In essence, the MSR solution allows both field aligned-flows and some 
cross-field flows (along $\mbf{\hat{b}_1}$). For the MSR solution to be compatible with 
the MHD momentum equation in 
the direction of the suppressed cross-field flows (along $\mbf{\hat{b}_2}$)
the dynamic flow pressure must balance external gas pressure gradients, 
magnetic tension, and gravity {\it in the absence} of actual MSR flows. 
As a result, equation (\ref{b2}) shows that the coplanar minimum structure 
approximation, although not incompatible with the MHD momentum equation, 
restricts its generality along $\mbf{\hat{b}_2}$. 

Let us now discuss the extent of the approximation $u_{\perp _z}=0$ and its apparent 
contradiction with magnetic flux emergence or submergence. Vertical cross-field flows 
take place along polarity reversal lines where $B_z=0$, so $u_{\perp _z} \ne 0$ by 
definition in this case. 
Such flows are necessary for the emergence or submergence of magnetic flux 
in the solar 
atmosphere and have been studied by several authors (see, e.g., Strous et al. 1996; 
van Driel-Gesztelyi, Malherbe, \& D\'{e}moulin 2000; Chae, Moon, \& Pevtsov 2004). 
However, it is important 
to consider two aspects of flux injection (emergence or submergence) 
through the solar 
photosphere and their relation to an inductive technique such as the MSR. These aspects 
are the length and time scales involved in flux injection. 

It is generally accepted that flux emergence takes place along the polarity reversal 
lines in emerging flux regions and in granular length scales, much smaller than the 
supergranular scales of the resulting active regions. Recent high-resolution observations 
(Bernasconi et al. 2002; Pariat et al. 2004, and others) 
complement our understanding and suggest that in granular scales 
the convective overshoot carries short undulatory emerging field lines to the 
surface. Some of these lines are convex ($\mbf{\Omega}$-shaped) and do not 
need resistive effects to emerge/submerge rapidly because the drainage of the 
dense, unmagnetized, photospheric plasma is effectively assisted by the 
field-line geometry. 
However, many magnetic field lines are concave ($\mbf{U}$-shaped) because this 
dense plasma keeps their horizontal central part anchored in the photosphere while 
the more vertical edges of the line emerge in the atmosphere. 
The submergence of such $\mbf{U}$-loops may also occur ideally, 
because of their geometry, but their complete emergence requires 
resistive instabilities such as the Rayleigh-Taylor instability (Parker 1966) which 
occurs at length scales of the order few $Mm$ (Pariat et al. 2004), the Kelvin-Helmholtz 
instability (Diver, Brown, \& Rust 1996; Bernasconi et al. 2002) at similar length 
scales, or a sudden reshaping caused by the intense magnetic tension force 
developed in these lines with small radii of curvature 
(e.g., Priest 1982). In essence, the average 
ratio $|u_{\perp _z}/u_{|| _z}|$ is probably large (of the order, or larger than, unity) 
in granular length scales, but it may be smaller than unity in supergranular scales 
since (i) flux injection is not observed in supergranular scales and (ii) all the 
instabilities proposed for resistive flux emergence require length scales of the order 
few $arcsec$, much smaller than supergranular scales. 
Macroscopic (supergranular) scales, however, are the scales 
of interest in inductive velocity calculation techniques. This is probably why our MSR 
velocity reasonably reproduces a number of macroscopic flow patterns in the 
active regions of \S5 but this this is also why it may perform poorly in small 
spatial scales. The pixel size of most modern vector magnetographs is of the order 
$1\arcsec$ or less ($0.55\arcsec$ for the IVM data used here) so the above-mentioned 
instability length scales and the subsequent small-scale dynamical activity 
are most probably resolved, at least partially. 

On the other hand, the emergence time scales for active regions are of the order few, 
to several, days. In an inductive velocity calculation the discretization of the 
induction equation (\ref{iz}) requires the smallest possible time differences between 
two successive vector magnetograms. If the Doppler velocity is used, 
however, time scales of at least a few tens of $min$ are necessary to average the 
Doppler signal and to eliminate high-frequency oscillatory effects. In balancing 
between the two requirements, an optimal time scale of the order $20-30\;min$ in 
measurements having a cadence of a few $min$ 
would both allow the discretized induction 
equation to work acceptably and the Doppler velocity to be properly averaged. 
This time scale is so much smaller than 
emerging flux region time scales that in practice one assumes that the evolution 
seen between two successive vector magnetograms is dominated by induction flows with only 
minor contribution from flux injection (emergence or submergence). 

From the above discussion it becomes rather clear that the ratio $|u_{\perp _z}/u_{|| _z}|$ 
in the length and time scales of interest to an inductive velocity calculation is either 
smaller or much smaller than unity, but it may become larger or much larger than unity 
in the small spatial and temporal scales at work in case of rapid flux emergence or 
submergence. This is why the MSR solution breaks down locally for, say, the small emerging 
pore in NOAA AR 8210 (\S5.2) or the canceling magnetic features reported by 
Chae, Moon, \& Pevtsov (2004). In the former case, the intense horizontal flows of the 
emerging structure are not reproduced. In the latter case, the significant 
downward $u_{\perp _z}$ cannot be reproduced by construction of the MSR solution. 
Furthermore, if these flux cancellations are caused by resistive 
effects, an ideal MHD technique is impossible to realistically capture 
the dynamics of the process. 

It is interesting to examine the feedback provided by 
numerical simulation results of ideal flux emergence 
(Fan 2001; Magara \& Longcope 2003; Magara 2004). 
In all studies, different geometries of the emerging magnetic field lines are 
simultaneously considered. In Fan (2001) a nearly horizontal line corresponding 
to the axis of the ascending flux tube never emerges fully from the photosphere 
although the upper part of the tube rises well above the photospheric plane. 
In Magara \& Longcope (2003) and Magara (2004) nearly horizontal field lines 
eventually reach substantial heights. It is argued, however, 
that the emergence behavior depends critically on 
both the geometry of the line and its assumed internal structure. Nearly horizontal 
lines imply a weaker plasma drainage toward their footpoints, so the gravitational 
force is more effective in blocking the expansion of the line. The drainage is more 
intense close to the 
footpoints of the line, where the magnetic field is more vertical, 
and hence these parts of the line ascend faster than the central part where a dip 
eventually develops and more unmagnetized plasma is stably accumulated. 
Magara \& Longcope (2003) and Magara (2004) 
argue that the mass accumulation in the dip eventually enhances the 
magnetic forces so that the line reshapes, probably reconnects, and eventually 
expands upward. In any case, the local apex(es) of any ascending or 
descending twisted magnetic field line 
are horizontal, so in these locations $u_{\perp _z} \ne 0$. In strongly curved 
field lines, however, these locations occupy small parts of the line so whether 
$u_{\perp _z} \ne 0$ might depend on the considered length scales, 
as discussed previously. The above modeling studies appear to show 
that the upward motion of a magnetic field line is slower if the line is horizontal 
and nearly untwisted as opposed to the faster upward motion of twisted magnetic field 
lines which are tangent to the photospheric plane only locally. In other words, these 
results may imply that the average 
$|u_{\perp _z}/u_{|| _z}| \gtrsim 1$ in small scales, near the local apexes of the line, 
while $|u_{\perp _z}/u_{|| _z}| < 1$ or $|u_{\perp _z}/u_{|| _z}| \ll 1$ if averaged over 
the entire length of the line.
\subsection{Concluding remarks}
All different inductive techniques must resort to an additional constraint 
in order to reach a unique solution of the under-determined ideal induction equation. 
The problem originates from our inability to measure the magnetic field vector in a 
succession of heights in and above the photosphere. The velocity field solution will be 
unique in the ideal limit of the induction equation only when simultaneous multi-height 
(i.e. photospheric and chromospheric) magnetic field measurements become available. 
The problem is also intractable when the resistive term of the 
induction equation is considered, because the magnetic diffusivity $\eta$ 
is unknown. Even if a reliable model of $\eta$ was used, however, new numerical 
techniques would be required to solve the resistive induction equation, so an  
envisioned step from ideal to non-ideal inductive techniques is inherently 
nontrivial. 

Some authors suggest that the 
additional constraint needed to uniquely determine the solution to the vertical 
component of the ideal induction equation is readily available and is the LCT 
velocity $\mbf{v_{LCT}}$ combined with the geometrical conjecture of 
D\'{e}moulin \& Berger (2003) (equation (\ref{db})). This additional information 
is fully exploited by the ILCT technique of Welsch et al. (2004). Nevertheless, 
we choose not to use $\mbf{v_{LCT}}$ as it is currently measured for a number of 
reasons: the map of $\mbf{v_{LCT}}$ is coarser than the input maps; $\mbf{v_{LCT}}$ 
is not really consistent with the induction equation; $\mbf{v_{LCT}}$ may incorporate 
HD flows of the field-free plasma; and $\mbf{v_{LCT}}$ represents only part of the 
kinetic power of the magnetized plasma in an AR, determined by a single choice of the 
LCT's FWHM apodizing window. As a result, $\mbf{v_{LCT}}$ is not unique. In the 
absence of a physical criterion for the selection of the above control parameter, 
it is not clear which prescription for $\mbf{v_{LCT}}$ should be chosen. 
We feel that if this central problem of the preferred LCT length scale 
is somehow addressed in the future the $\mbf{v_{LCT}}$-solution 
together with equation (\ref{db}) 
might indeed be useful in constraining the induction equation.  

Comparing the MSR results with the results of the ILCT and the MEF techniques for NOAA 
AR 8210, we find that (i) the MSR provides a rather more detailed velocity field 
solution, assisted also by the field-aligned velocity calculation, and (ii) the MSR 
velocity tends to be larger than the ILCT and MEF velocities. This may be 
because the assumption $u_{\perp _z}=0$ in the MSR solution introduces additional 
horizontal flows (\S6.1) and/or 
because the ILCT and MEF solutions rely on smoothed (from LCT) 
and explicitly minimal flows, respectively. This is an unclear point and needs to 
be addressed in the future. Qualitatively, the MSR velocity solution 
reasonably reproduces the major flow features, as seen in the magnetogram movies, for 
all three examples discussed in \S5. However, the MSR velocity field 
{\it does not} reproduce localized flows of emerging magnetic structures. 
There is no feedback from the MEF solution on this aspect, but the ILCT solution 
captures this feature since it relies on 
the LCT velocity which attributes a nonzero horizontal 
velocity to purely vertical flows. This discrepancy between the MSR and the 
ILCT results might be expected because the MSR method (and, apparently, the MEF 
method) relies only on the temporal variation of $B_z$ and it cannot track down a  
particular feature, whereas the ILCT method exhibits a ``memory'' in identifying 
and tracking down individual magnetic features and interprets their observed 
displacements by means of an assigned velocity. Moreover, the main approximation 
$u_{\perp _z}=0$ of the MSR approach is expected to break down locally in areas of 
intense, rapid flux emergence or submergence. 

Let us also underline that none of the existing inductive techniques can 
fully advance a given vector magnetogram in time to provide a short-term prediction 
for the evolution of the magnetic field vector based on the inferred inductive 
velocity. This would be useful for providing a continuous, or nearly continuous, 
boundary condition for data-driven 3D MHD models of the active-region corona. 
From single-height magnetic field measurements, however, 
only the vertical magnetic field can be advanced. 
Multi-height magnetic field measurements are absolutely necessary for 
advancing the horizontal magnetic field components. 

Until recently, only the various tracking 
algorithms were available to assess the velocity 
field in solar active regions. There are now four different techniques, namely  
Kusano's et al. (2002) method, the ILCT method of Welsch et al. (2004), the MEF 
method of Longcope (2004), and the MSR method of this study, that use the ideal 
induction equation in an attempt to advance our understanding of the magnetized plasma 
flows. A comparative evaluation of all four techniques is underway and 
aims to realize the overall benefit 
furnished by the various methodologies. One envisions a consensus on 
an optimal velocity field calculation 
that will decisively improve our understanding of magnetized 
flows in solar active regions. A such scheme could incorporate 
elements from the existing techniques and should be open to future 
physically-founded ideas. These ideas might be incorporated into the solution of the 
induction equation by means of our general solving methodology. 
\acknowledgements
We wish to thank an anonymous referee for insightful comments and suggestions. 
Data used here from Mees Solar Observatory, University of   
Hawaii, are produced with the support of NASA and AFRL grants.
This work has received partial support by NSF Grant ATM-0208104 
and by NASA Grant NAG5 13504.  
\clearpage

\clearpage
\newpage
\centerline{\includegraphics[width=19.cm,height=8.cm,angle=0]{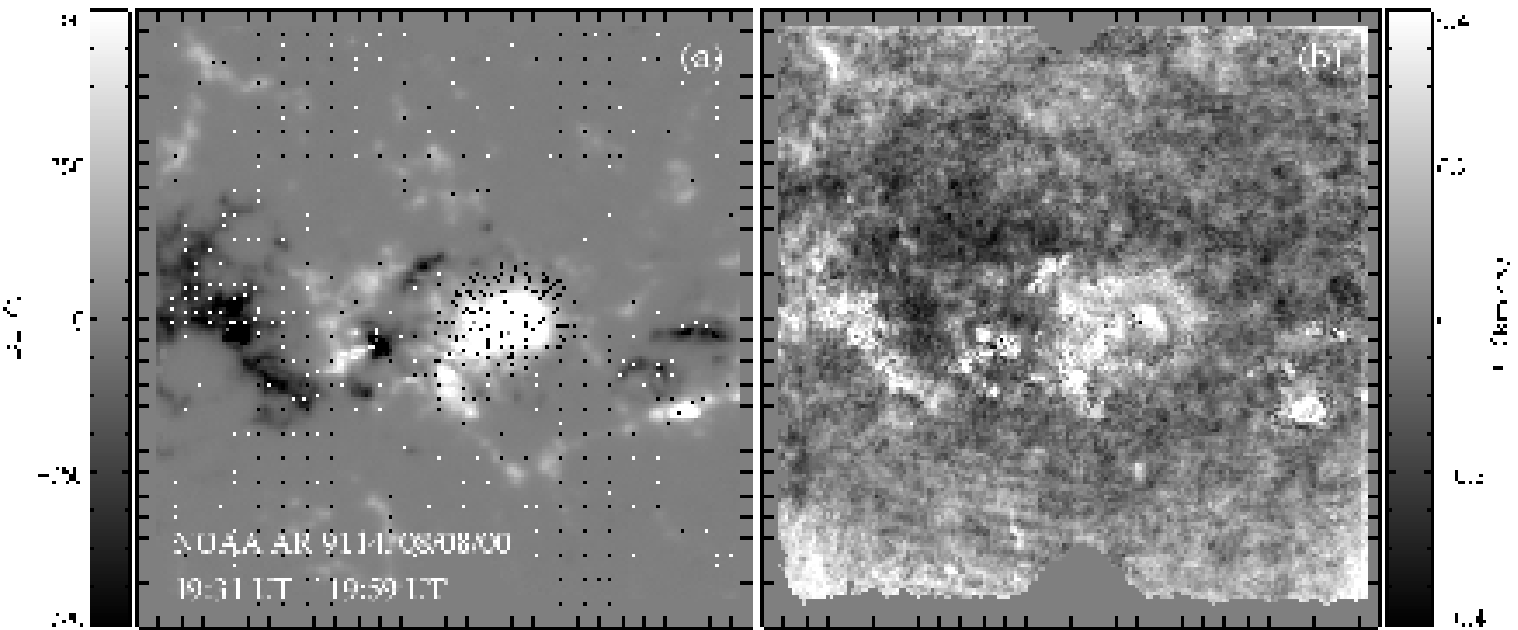}}
\figcaption{Average magnetic field and Doppler velocity in NOAA AR 9114 
for a pair of vector magnetograms observed by the IVM on 2000 August 8 at 
19:31 UT and at 19:59 UT, respectively. (a) Ambiguity-free average 
heliographic magnetic field vector on the image plane. 
A vector length equal to the tick mark separation 
corresponds to a horizontal magnetic field equal to $890\;G$. (b) Average 
Doppler velocity in the AR. Tick mark separation in both images is $10\arcsec$.
\label{f1}}
\newpage
\centerline{\includegraphics[width=15.cm,height=12.cm,angle=0]{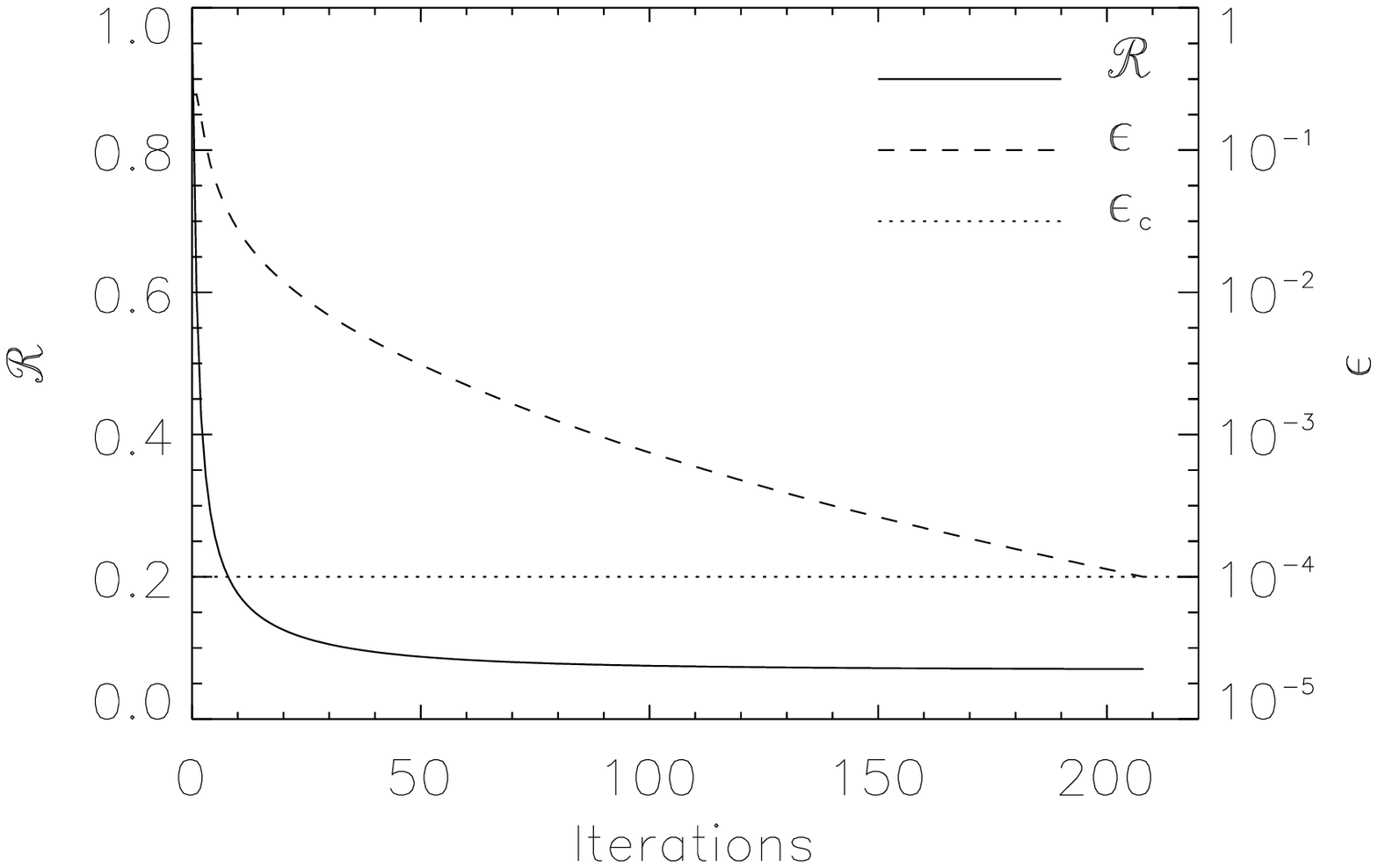}}
\figcaption{The convergence process in calculating the electrostatic field $\mbf{G}$. 
We show the minimization of the dimensionless ratio 
$\mathcal{R}$ (equation (\ref{c1}); solid line) 
and the fractional tolerance limit $\varepsilon$ 
that controls the minimization process (equation (\ref{epsi}); dashed line). 
The iterations stop when 
$\varepsilon$ becomes smaller that a prescribed fractional tolerance limit 
$\varepsilon _c$, set to $10^{-4}$ (dotted line). 
\label{f2}}
\newpage
\centerline{\includegraphics[width=19.cm,height=8.cm,angle=0]{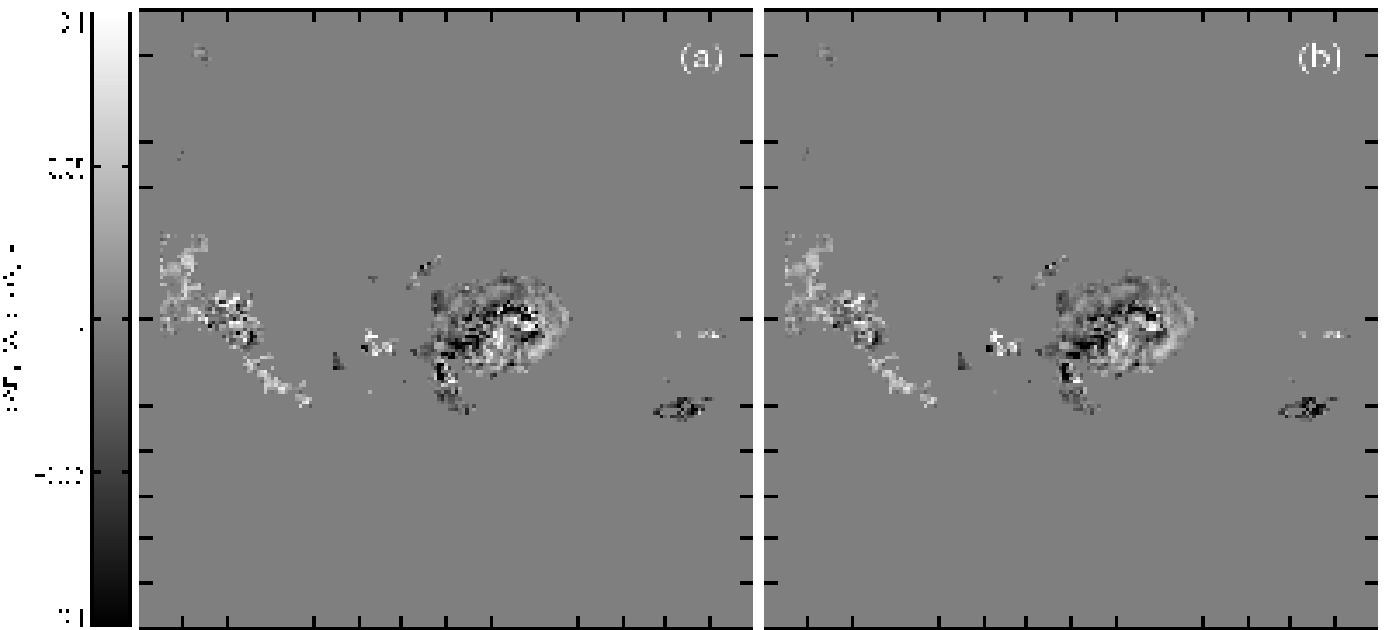}}
\centerline{\includegraphics[width=8.cm,height=6.cm,angle=0]{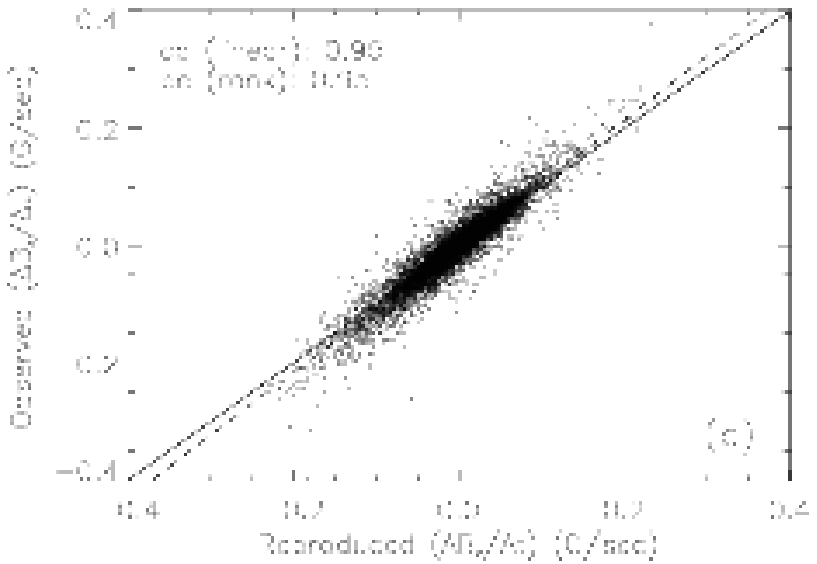}}
\figcaption{The temporal variation of the vertical magnetic field for the pair 
of magnetograms taken at 19:31 UT and at 19:59 UT in NOAA AR 9114: 
(a) Observed temporal variation. 
(b) Reproduced temporal variation, using the MSR velocity solution in the induction 
equation to solve for the vertical magnetic field (see text for details). 
Information is shown only for areas of strong magnetic fields. Tick mark 
separation in both images is $20\arcsec$. 
(c) Comparison between the observed and the reproduced temporal variations. 
The dashed line corresponds to the best linear fit between the two 
quantities, while the analytical relation of equality 
is shown by the solid line. The linear and the Spearman rank 
correlation coefficients (cc) are also indicated.
\label{f3}}
\newpage
\centerline{\includegraphics[width=19.cm,height=8.cm,angle=0]{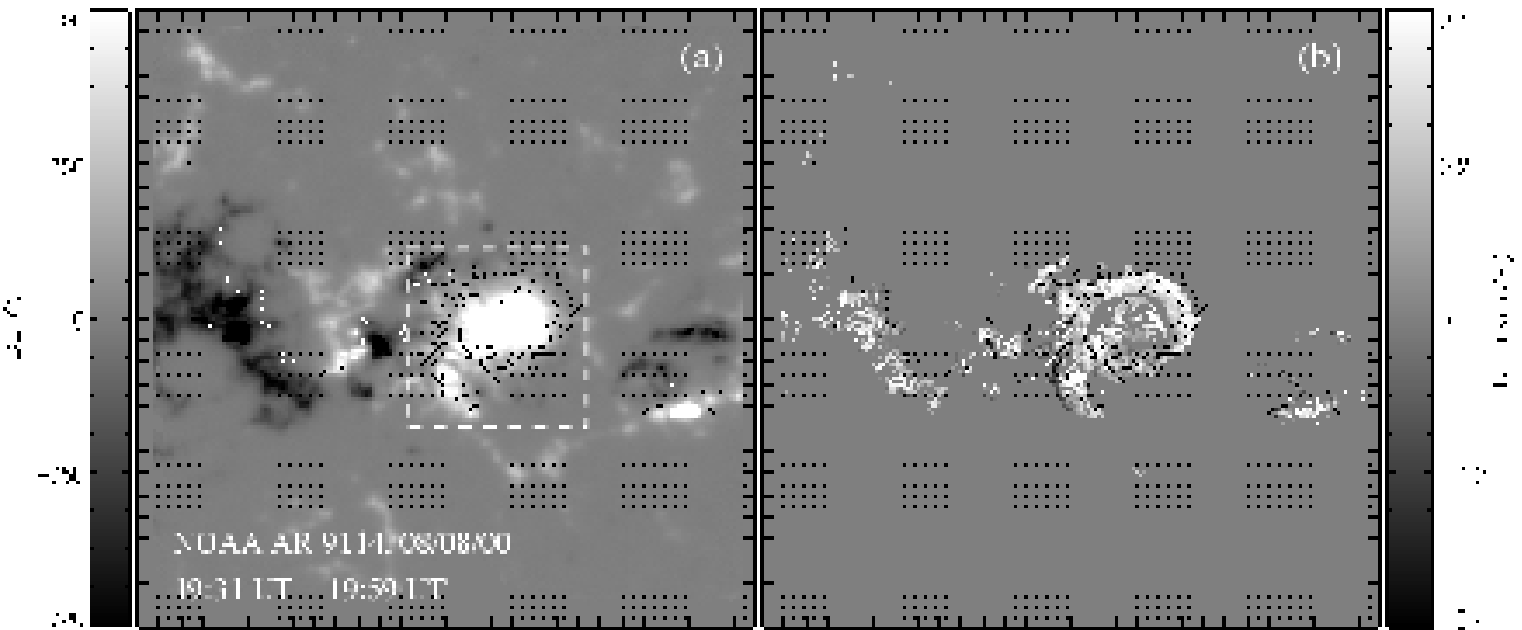}}
\figcaption{The MSR velocity field solution for the pair of magnetograms obtained 
at 19:31 UT and at 19:59 UT in NOAA AR 9114: (a) The horizontal velocity solution 
plotted on top of the average vertical magnetic field. The dashed rectangle encloses 
a rotating sunspot with distinctive outflows (see text for details). 
(b) The horizontal velocity solution plotted on top of the calculated vertical velocity. 
For both images, a vector length equal to the tick mark separation corresponds to a 
horizontal velocity field approximately equal to $1.5\;km\;s^{-1}$.
Tick mark separation is $10\arcsec$. 
\label{f4}}
\newpage
\centerline{\includegraphics[width=19.cm,height=11.cm,angle=0]{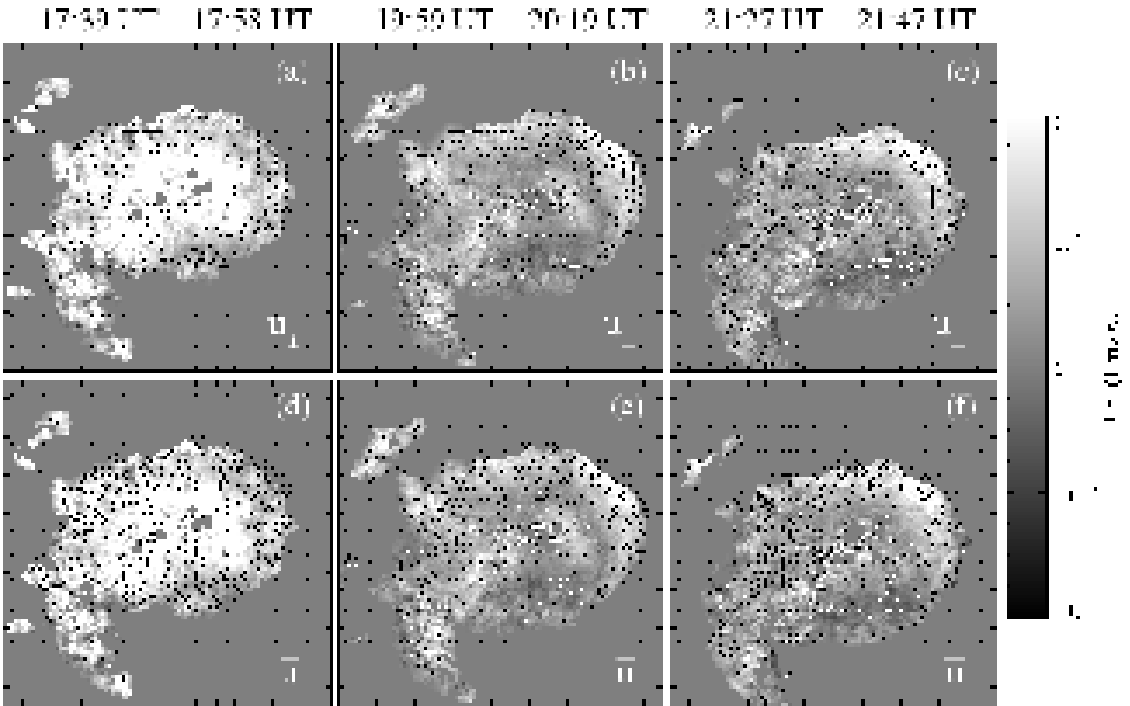}}
\figcaption{The MSR velocity field solution for a sub-region of NOAA AR 9114 
indicated by the dashed rectangle in Figure \ref{f4}a. Three different pairs 
of the AR vector magnetograms are used, obtained by the IVM on 2000 August 8. 
The times of each magnetogram pair are 
indicated at the top. The upper row of images (a, b, c) provides the horizontal 
cross-field velocity vector while the 
lower row of images (d, e, f) provides the total horizontal 
velocity vector, both plotted on top of the calculated vertical velocity. 
Each column of images (a,d; b,e; c,f) corresponds to the same pair of vector 
magnetograms. A vector length equal to the tick 
mark separation corresponds to a horizontal velocity field approximately equal 
to $1\;km\;s^{-1}$ in all images. Tick mark separation in all images is $5\arcsec$.
\label{f5}}
\newpage
\centerline{\includegraphics[width=19.cm,height=8.cm,angle=0]{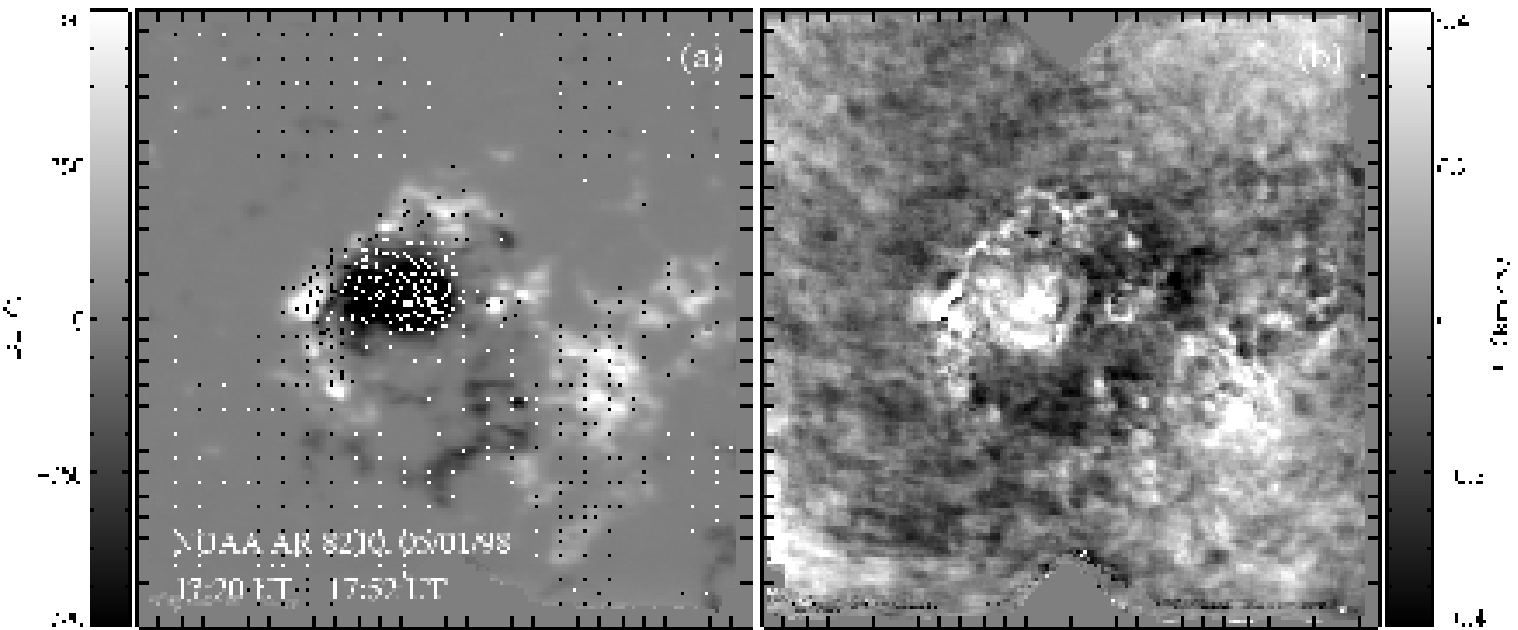}}
\figcaption{Average magnetic field and Doppler velocity in NOAA AR 8210
for a pair of vector magnetograms observed by the IVM on 1998 May 1 at 
17:20 UT and at 17:52 UT, respectively. (a) Ambiguity-free average 
heliographic magnetic field vector on the image plane. 
A vector length equal to the tick mark separation 
corresponds to a horizontal magnetic field equal to $980\;G$. (b) Average 
Doppler velocity in the AR. Tick mark separation in both images is $10\arcsec$.
\label{f6}}
\newpage
\centerline{\includegraphics[width=19.cm,height=8.cm,angle=0]{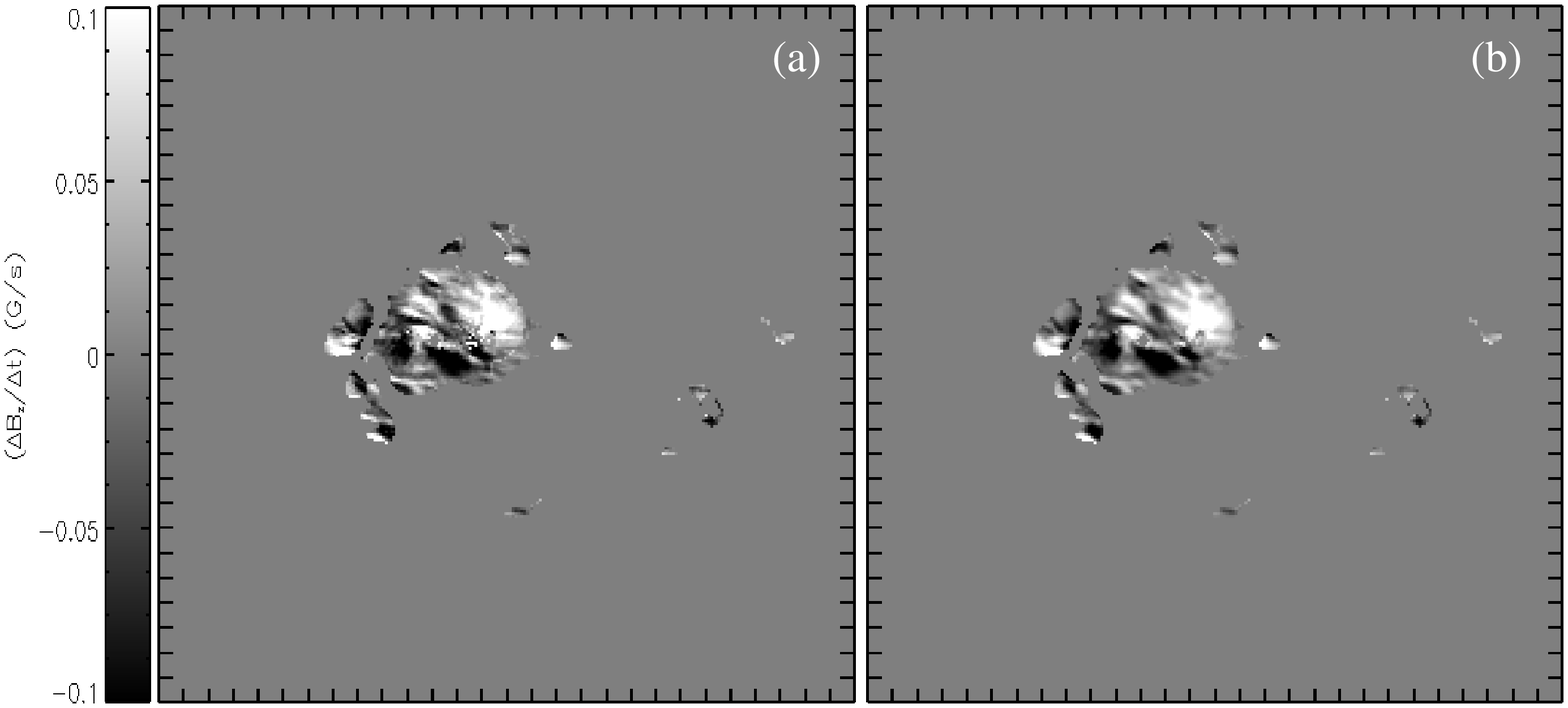}}
\centerline{\includegraphics[width=8.cm,height=6.cm,angle=0]{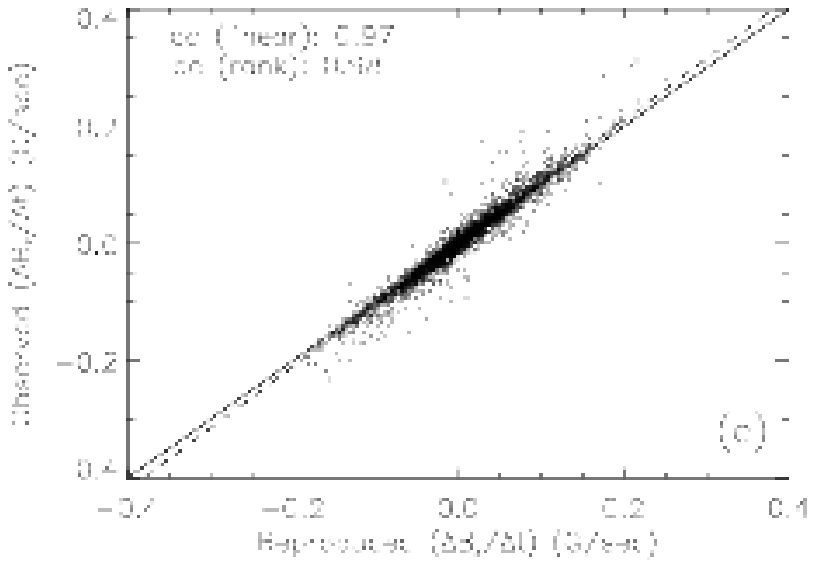}}
\figcaption{The temporal variation of the vertical magnetic field for the pair 
of magnetograms obtained at 17:20 UT and at 17:52 UT in NOAA AR 8210: 
(a) Observed temporal variation. 
(b) Reproduced temporal variation, using the MSR velocity solution in the induction 
equation to solve for the vertical magnetic field. 
Information is shown only for areas of strong magnetic fields. Tick mark 
separation in both images is $10\arcsec$. 
(c) Comparison between the observed and the reproduced temporal variations. 
The dashed line corresponds to the best linear fit between the two 
quantities, while the analytical relation of equality 
is shown by the solid line. The linear and the Spearman rank 
correlation coefficients (cc) are also indicated.
\label{f7}}
\newpage
\centerline{\includegraphics[width=17.cm,height=15.cm,angle=0]{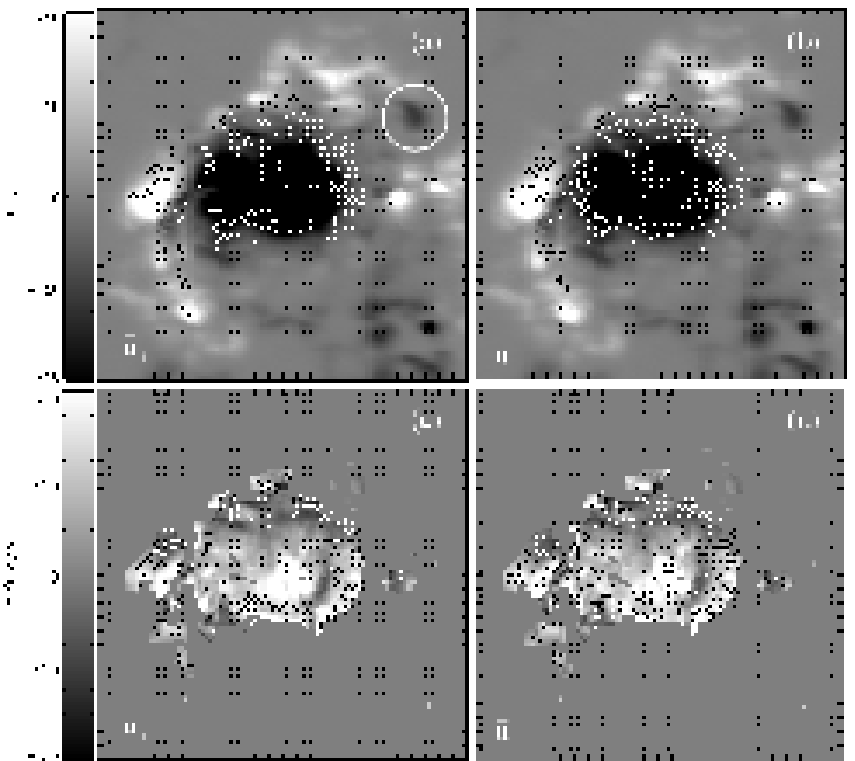}}
\figcaption{The MSR velocity field solution for the $\delta$-sunspot in NOAA AR 
8210, calculated for the pair of magnetograms obtained at 17:20 UT and at 17:52 UT.
Only the calculated velocity for the strong-field regions is shown. The grayscale 
background in the upper row of images is the average vertical magnetic field, 
while for the lower row of images it is the vertical MSR velocity. The vector 
field in the left column of images corresponds to the cross-field MSR velocity 
$\mbf{u_{\perp}}$ while in the right column of images it corresponds to 
the total MSR velocity $\mbf{u}$. In all images, tick mark separation is $5\arcsec$. 
A vector length equal to the tick mark separation corresponds to 
a horizontal velocity of $\sim 1.15\;km\;s^{-1}$. 
The white oval in (a) indicates a small emerging flux sub-region discussed in the text. 
\label{f8}}
\newpage
\centerline{\includegraphics[width=18.cm,height=11.cm,angle=0]{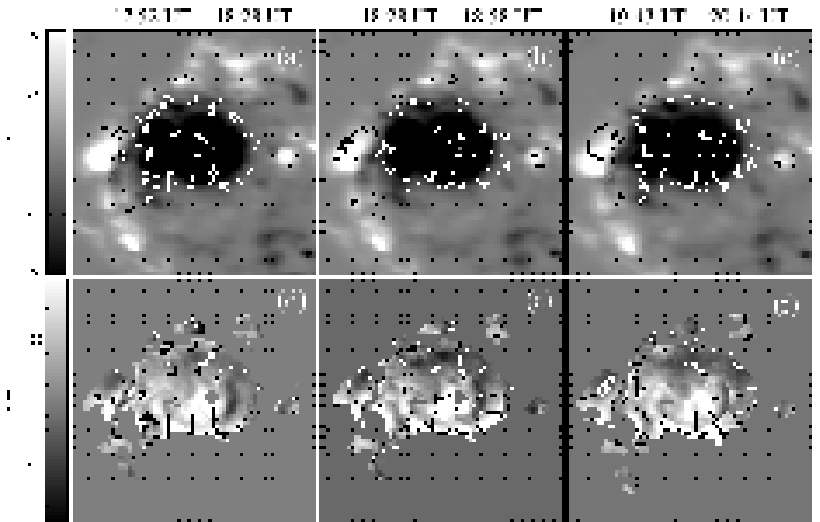}}
\figcaption{The MSR velocity field solution for the $\delta$-sunspot in NOAA AR 
8210 calculated for three different pairs of vector magnetograms obtained by the 
IVM on 1998 May 1. The vector field in each image corresponds to the total 
MSR velocity $\mbf{u}$. Each column of images corresponds to the same pair 
of magnetograms. The UT times of each pair are given at the top of each column 
of images. 
The grayscale background in the upper row of images corresponds to the 
average vertical magnetic field for each pair, while in the lower row of images 
it is the MSR vertical velocity. In all images, tick mark separation is $5\arcsec$. 
A vector length equal to the tick mark separation corresponds to a horizontal 
velocity of $\sim 1.3\;km\;s^{-1}$. 
\label{f9}}
\newpage
\centerline{\includegraphics[width=19.cm,height=8.cm,angle=0]{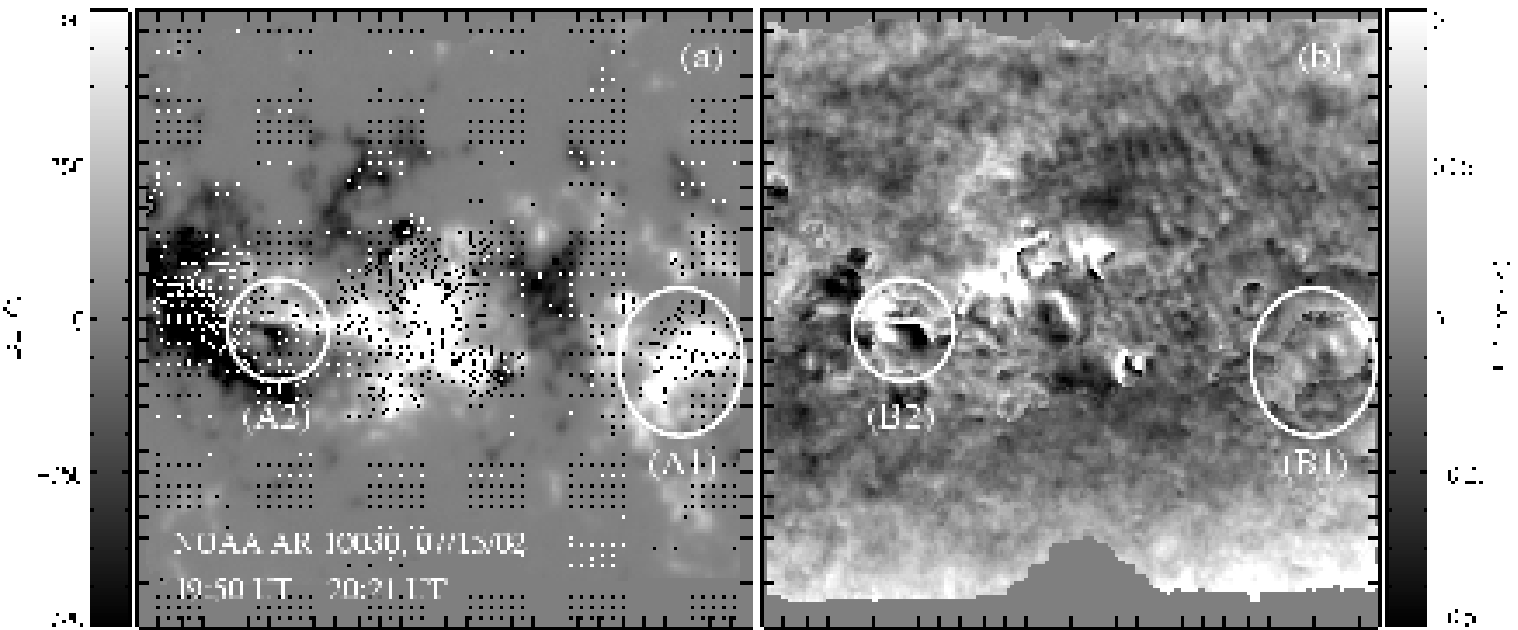}}
\figcaption{Average magnetic field and Doppler velocity in NOAA AR 10030 
for a pair of vector magnetograms observed by the IVM on 2002 July 15 at 
19:50 UT and at 20:21 UT, respectively. (a) Ambiguity-free average 
heliographic magnetic field vector on the image plane. 
A vector length equal to the tick mark separation 
corresponds to a horizontal magnetic field equal to $1250\;G$. (b) Average 
Doppler velocity in the AR. The white ovals correspond to two sub-regions of 
the AR discussed in the text. Tick mark separation in both images is $10\arcsec$.
\label{f10}}
\newpage
\centerline{\includegraphics[width=19.cm,height=8.cm,angle=0]{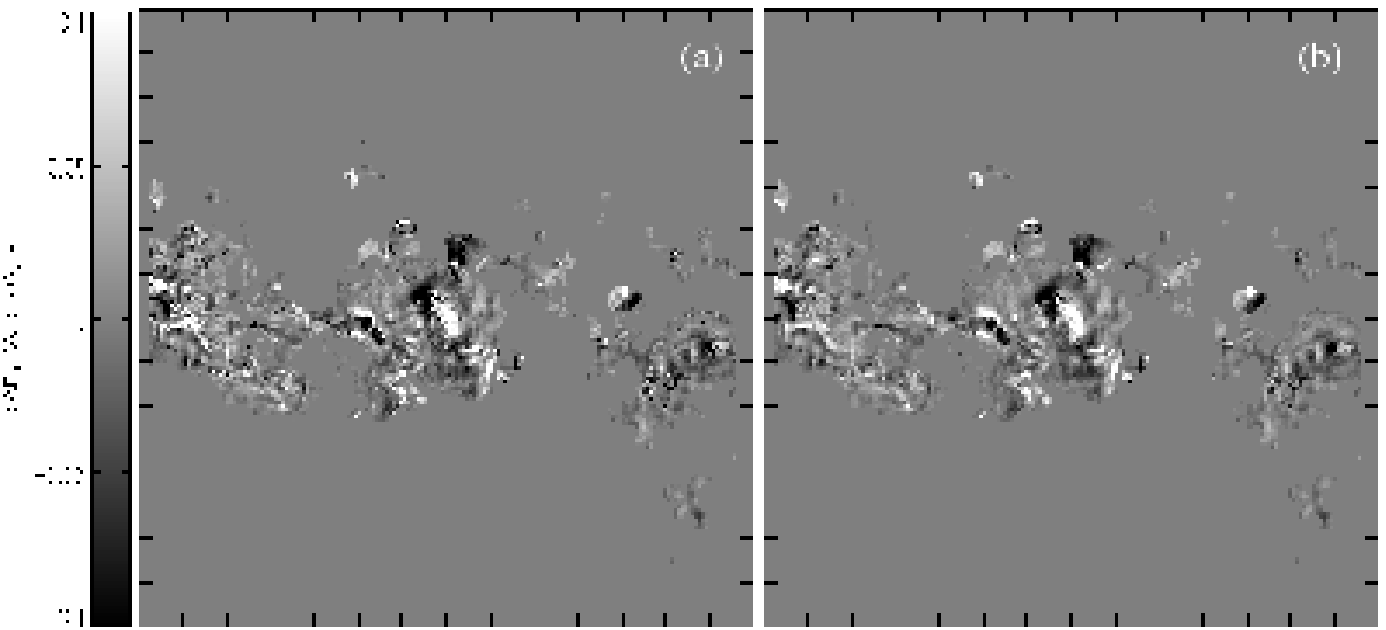}}
\centerline{\includegraphics[width=8.cm,height=6.cm,angle=0]{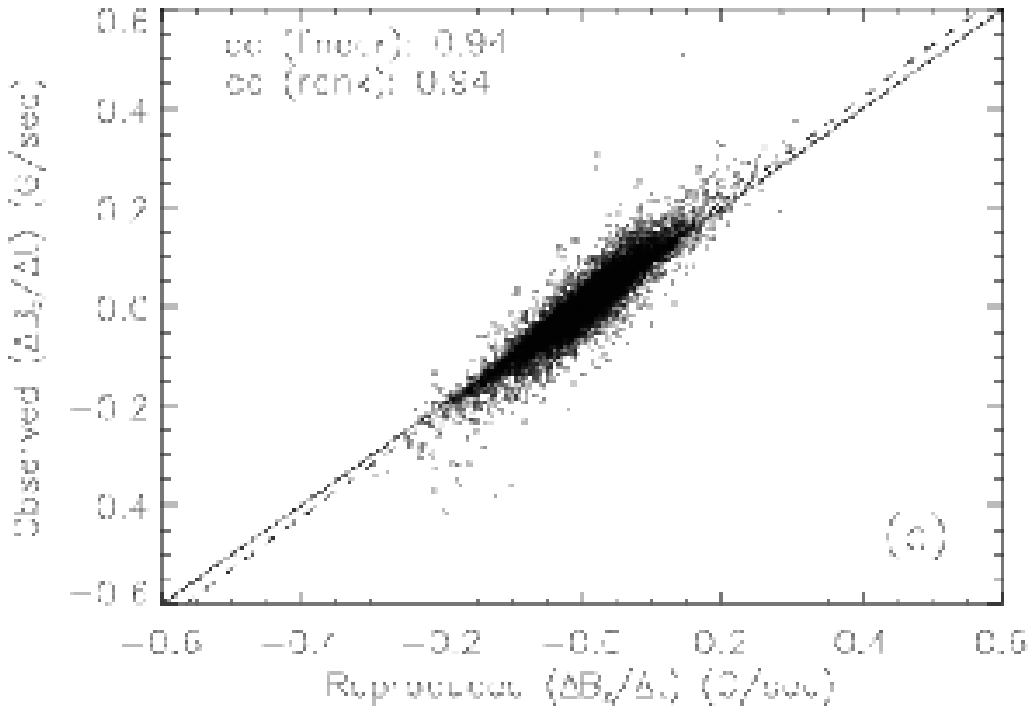}}
\figcaption{The temporal variation of the vertical magnetic field for the pair 
of magnetograms obtained at 19:50 UT and at 20:21 UT in NOAA AR 10030: 
(a) Observed temporal variation. 
(b) Reproduced temporal variation, using the MSR velocity solution in the induction 
equation to solve for the vertical magnetic field. 
Information is shown only for areas of strong magnetic fields. Tick mark 
separation in both images is $20\arcsec$. 
(c) Comparison between the observed and the reproduced temporal variations. 
The dashed line corresponds to the best linear fit between the two 
quantities, while the analytical relation of equality 
is shown by the solid line. The linear and the Spearman rank 
correlation coefficients (cc) are also indicated. 
\label{f11}}
\newpage
\centerline{\includegraphics[width=12.cm,height=15.cm,angle=0]{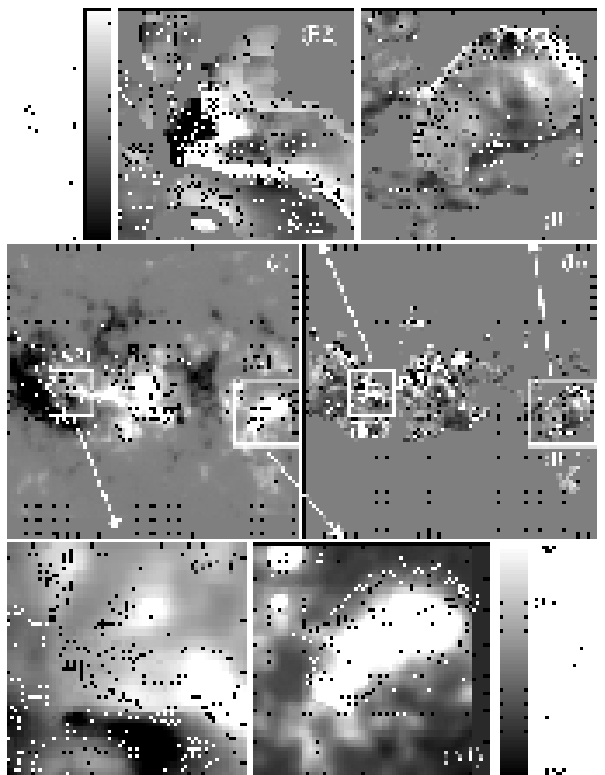}}
\figcaption{The MSR velocity field solution for the pair of vector magnetograms 
of NOAA AR 10030 obtained 
at 19:50 UT and at 20:21 UT. Only the velocity solution for the strong magnetic 
field regions is shown. 
(a) The horizontal velocity solution 
plotted on top of the average vertical magnetic field. 
(b) The horizontal velocity solution plotted on top of the calculated vertical velocity. 
Tick mark separation in (a) and (b) is $10\arcsec$. A vector length equal to the 
tick mark separation in (a) and (b) corresponds to a horizontal velocity field of 
$\sim 1\;km\;s^{-1}$. Two details of the velocity solution have been magnified in 
(a) (details (A1) and (A2)) and (b) (details (B1) and (B2)). 
Tick mark separation in the details is $2\arcsec$. A vector length equal to the 
tick mark separation in the details corresponds to a horizontal velocity field 
of $\sim 2\;km\;s^{-1}$. 
\label{f12}}
\newpage
\centerline{\includegraphics[width=15.cm,height=13.cm,angle=0]{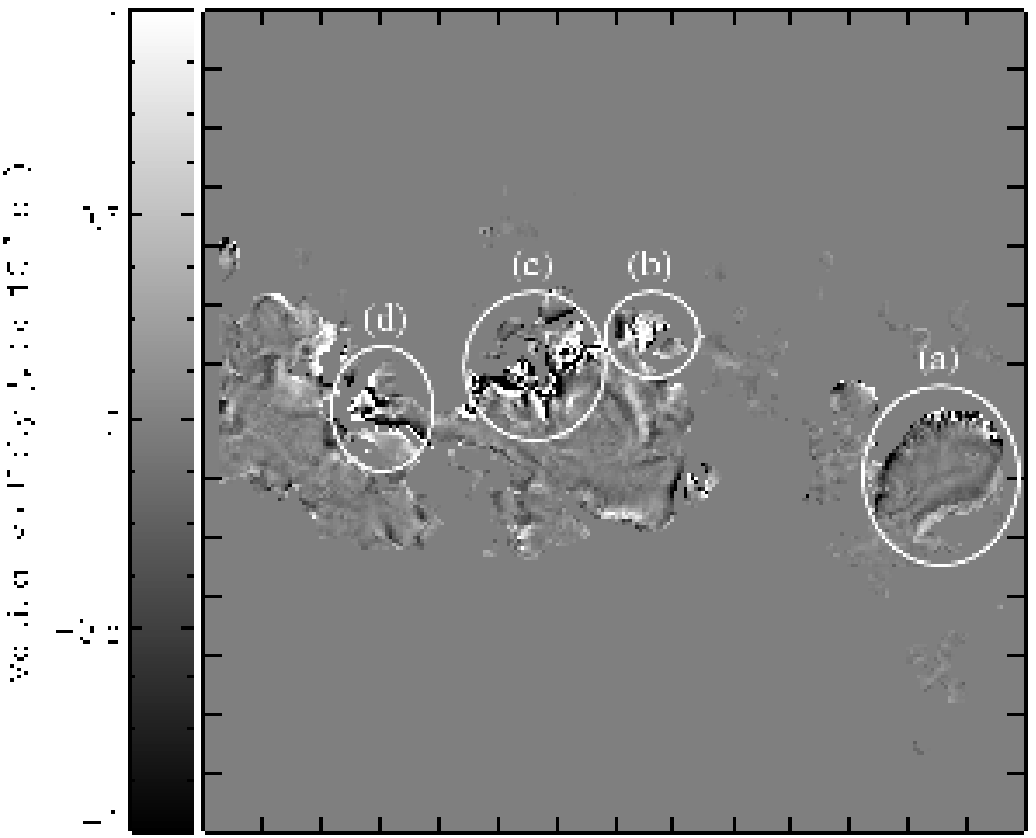}}
\figcaption{The vertical vorticity in NOAA AR 10030, calculated using the MSR 
velocity field solution for areas of strong magnetic field. 
Four extended areas of strong vorticity have been indicated by 
the ovals (a), (b), (c), and (d). Tick mark separation is $20\arcsec$. 
\label{f13}}

\begin{thebibliography}{99}
\bibitem[]{101} Abbett, W. P., AGU Fall Meeting 2003, Abstract \#SM11A-03
\bibitem[]{102} Alissandrakis, C. E., 1981, A\&A, 100, 197
\bibitem[]{103} Amari, T., Boulmezaoud, T. Z., \& Miki\'{c}, Z., 1999, A\&A, 350, 1051
\bibitem[]{104} Amari, T. Luciani, J. F., Aly, J. J., Miki\'{c}, Z., \& Linker, J., 
2003, ApJ, 595, 1231
\bibitem[]{105} Berger, M. A., 1999, Plasma Phys. Contr. Fusion, 41, 167
\bibitem[]{106} Berger, M. A., \& Field, G. B., 1984, Journal Fluid Mech., 147, 133
\bibitem[]{143} Berger, Th. E., L\"{o}fdahl, M. G., Shine, R. S., \& Title, A. M., 
1998, ApJ, 495, 973
\bibitem[]{1005} Bernasconi, P. N., Rust, D. M., Georgoulis, M. K., \& LaBonte, B. J., 
2002, SoPh, 209, 119
\bibitem[]{131} Boris, J. P., Landsberg, A. M., Oran, E. S., \& Gardner, J. H., 
LCPFCT - Flux Corrected Transport Algorithm for Solving Generalized Continuity Equations, 
NRL/MR/6410-93-7192, p.23
\bibitem[]{200} Brown, D. S., Nightingale, R. W., Alexander, D., Schrijver, C. J., 
Metcalf, Th. R., Shine, R. A., Title, A. M., \& Wolfson, C. J., 2003, SoPh, 216, 79
\bibitem[]{108} Chae, J., 2001, ApJ, 560, L95
\bibitem[]{996} Chae, J., Moon, Y.-J., \& Pevtsov, A. A., 2004, ApJ, 602, L65
\bibitem[]{109} D\'{e}moulin, P., \& Berger, M. A., 2003, SoPh, 215, 203
\bibitem[]{1006} Diver, D. A., Brown, J. C., \& Rust, D. M., 1996, SoPh, 168, 105
\bibitem[]{1002} van Driel-Gesztelyi, L., Malherbe, J.-M., \& D\'{e}moulin, P., 
2000, A\&A, 364, 845
\bibitem[]{990} Fan, Y., 2001, ApJ, 554, L111
\bibitem[]{1007} Fisher, G. H., AGU Fall Meeting 2003, Abstract \#SH51A-01
\bibitem[]{110} Gary, G. A., \& Hagyard, M. J., 1990, SoPh, 126, 21
\bibitem[]{111} Gary, G. A., \& Moore, R. L., 2004, ApJ, 611, 545
\bibitem[]{112} Georgoulis, M. K., \& LaBonte, B. J., 2004, ApJ, 615, 1029
\bibitem[]{113} Georgoulis, M. K., LaBonte, B. J., \& Metcalf, Th. R., 
2004, ApJ, 602, 446
\bibitem[]{1001} Georgoulis, M. K., Rust, D. M., Bernasconi, P. N., \& Schmieder, 
B., 2002, ApJ, 575, 506
\bibitem[]{114} Gudiksen, B. V., \& Nordlund, A., 2002, ApJ, 572, L113 
\bibitem[]{115} Harvey, K., \& Harvey, J., 1973, SoPh, 28, 61 
\bibitem[]{997} Jackson, J. D., 1962, Classical Electrodynamics (New York: Wiley)
\bibitem[]{116} Kosovichev, A. G., 2002, AN 323, No. 3/4, 186
\bibitem[]{117} Kusano, K., Maeshiro, T., Yokoyama, T., \& Sakurai, T., 2002, ApJ, 577, 
501
\bibitem[]{118} Landolfi, M., \& degl'Innocenti, E. L., 1982, SoPh, 78, 355
\bibitem[]{119} Lionello, R., Velli, M., Einaudi, G., \& Miki\'{c}, Z., 1998, ApJ, 494, 
840
\bibitem[]{120} Liu, Y., Jiang, Y., Ji, H., Zhang, H., \& Wang, H., 2003, ApJ, 593, L137
\bibitem[]{121} Longcope, D. W., 2004, ApJ, 612, 1181
\bibitem[]{122} L\'{o}pez Fuentes, M. C., D\'{e}moulin, P., Mandrini, C. H., 
Pevtsov, A. A., \& van Driel-Gesztelyi, L., 2003, A\&A, 397, 305
\bibitem[]{991} Magara, T., 2004, ApJ, 605, 480
\bibitem[]{992} Magara, T., \& Longcope, D. W., 2003, ApJ, 586, 630
\bibitem[]{123} Metcalf, Th. R., Jiao, L., McClymont, A. N., Canfield, R. C., \& 
Uitenbroek, H., 1995, ApJ, 439, 474
\bibitem[]{124} Meytlis, V. P., \& Strauss, H. R., 1993, SoPh, 145, 111
\bibitem[]{125} Mickey, D. L., Canfield, R. C., LaBonte, B. J., Leka, K. D., 
Waterson, M. F., \& Weber, H. M., 1996, SoPh, 168, 229
\bibitem[]{126} Moon, Y.-J., Chae, J., Wang, H., Choe, G. S., \& Park, Y. D., 2002, 
ApJ, 580, 528
\bibitem[]{127} Nindos, A., \& Zhang, H., 2002, ApJ, 573, L133 
\bibitem[]{129} Nindos, A., \& Zirin, H., 1998, SoPh, 182, 381
\bibitem[]{130} November, L. J., \& Simon, G. W., 1988, 333, 427
\bibitem[]{994} Pariat, E., Aulanier, G., Schmieder, B., Georgoulis, M. K., 
Rust, D. M., \& Bernasconi, P. N., 2004, ApJ, 614, 1099
\bibitem[]{1003} Parker, E. N., 1966, ApJ, 145, 811
\bibitem[]{132} Pohjolainen, S., Maia, D., Pick, M., Vilmer, N., Khan, J. I., 
Otruba, W., Warmuth, A., Benz, A., Alissandrakis, C., \& Thompson, B. J., 2001, ApJ, 
566, 421 
\bibitem[]{133} Press, W. H., Flannery, B. P., Teukolsky, S. A., \& Vetterling, W. T., 
1992, Numerical Recipes: The Art of Scientific Computing (New York: Cambridge Univ. 
Press), pp. 655 - 667
\bibitem[]{1004} Priest, E. R., 1982, Solar Magnetohydrodynamics, 
(Dordrecht: Reidel Publishing Company), p. 102
\bibitem[]{201} Potts, H. E., Barrett, R. K., \& Diver, D. A., 2004, A\&A, 424, 253
\bibitem[]{134} Romano, P., Contarino, L., \& Zuccarello, F., 2003, SoPh, 218, 137
\bibitem[]{995} Roudier, Th., Rieutord, M., Malherbe, J. M., \& Vigneau, J., 1999, 
A\&A, 349, 301 
\bibitem[]{135} Roussev, I. I., Sokolov, I. V., Forbes, T. G., Gombosi, T. I., 
Lee, M. A., \& Sakai, J. I., 2004, ApJ, 605, L73
\bibitem[]{136} Schmidt, H. U., 1964, in AAS-NASA Symposium on the Physics of Solar 
Flares, ed. W. N. Hess (NASA SP-50), 107
\bibitem[]{137} Sterling, A. C., \& Moore, R. L., 2001, JGR, 106, A11, 25227
\bibitem[]{138} Strous, L. H., Scharmer, G., Tarbell, T. D., Title, A. M., \& 
Zwaan, C., 1996, A\&A, 306, 947
\bibitem[]{139} Thompson, B. J., Cliver, E. W., Nitta, N., Delan\'{e}e, C., \& 
Delaboudini\'{e}re, J.-P., 2000, GRL, 27, 1431 
\bibitem[]{140} Wang, T., Yan, Y., Wang, J., Kurokawa, H., \& Shibata, K., 2002, 
ApJ, 572, 580
\bibitem[]{141} Warmuth, A., Hanslmeier, A., Messerotti, A., Moretti, P.-F., \& 
Otruba, W., 2000, SoPh, 194, 103
\bibitem[]{142} Welsch, B. T., Fisher, G. H., Abbett, W. P., \& Regnier, S., 2004, 
ApJ, 610, 1148
\end{thebibliography}
\end{document}